\documentclass[aps,pra,twocolumn,superscriptaddress,notitlepage,nofootinbib,longbibliography]{revtex4-2}
\usepackage{mathrsfs}
\usepackage{epsfig}
\usepackage{graphicx}
\usepackage{amsfonts}
\usepackage[figuresright]{rotating}
\usepackage{amssymb}
\usepackage{amsmath}
\usepackage{dcolumn}
\usepackage{bm}
\usepackage{xcolor}
\usepackage{color}
\usepackage{braket}
\usepackage{units}
\usepackage{xspace}

\usepackage{bbold}
\definecolor{mypink3}{cmyk}{0, 0.7808, 0.4429, 0.1412}
\definecolor{mypink1}{rgb}{0.858, 0.188, 0.478}
\definecolor{mypink2}{RGB}{219, 48, 122}
\usepackage[colorlinks=true, allcolors=mypink1]{hyperref}
\usepackage[fleqn]{mathtools}

\usepackage[shortlabels]{enumitem}
\newcommand{\ECNU}{Quantum Institute for Light and Atoms, State Key Laboratory of Precision Spectroscopy, Department of Physics, School of Physics and Electronic Science, East China Normal University, Shanghai 200062, China}
\newcommand{\SBH}{Shanghai Branch, Hefei National Laboratory, Shanghai 201315, China}
\newcommand{\SPA}{School of Physics and Astronomy, and Tsung-Dao Lee Institute, Shanghai Jiao Tong University, Shanghai 200240, China}
\newcommand{\SRC}{Shanghai Research Center for Quantum Sciences, Shanghai 201315, China}
\newcommand{\CIC}{Collaborative Innovation Center of Extreme Optics, Shanxi University, Taiyuan, Shanxi 030006, China}
\begin{document}
\title{Estimation theory of photon-magnon coupling strength in a driven-dissipative double-cavity-magnon system}

\author{Jia-Xin Peng}
\affiliation{\ECNU}
\affiliation{\SBH}
\author{Baiqiang Zhu}
\affiliation{\ECNU}
\affiliation{\SBH}
\author{Weiping Zhang}
\affiliation{\SBH}
\affiliation{\SPA}
\affiliation{\SRC}
\affiliation{\CIC}
\author{Keye Zhang}
\email{kyzhang@phy.ecnu.edu.cn}
\affiliation{\ECNU}
\affiliation{\SBH}
\date{\today}

\begin{abstract}
Cavity-magnon systems are emerging as a fruitful architecture for the integration of quantum technologies and spintronic technologies, where magnons are coupled to microwave photons via the magnetic-dipole interaction. Controllable the photon-magnon (P-M) couplings provide a powerful means of accessing and manipulating quantum states in such hybrid systems. Thus determining the relevant P-M couplings is a fundamental task. Here we address the quantum estimation problem for the P-M coupling strength in a double-cavity-magnon system with drive and dissipation. The effects of various physical factors on the estimation precision are investigated and the underlying physical mechanisms are discussed in detail. Considering that in practical experiments it is almost infeasible to perform measurements on the global quantum state of this composite system, we identify the optimal subsystem for performing measurements and estimations. Further, we evaluate the performance of different Gaussian measurements, indicating that optimal Gaussian measurement almost saturates the ultimate theoretical bound on the estimation precision given by the quantum Fisher information.
\end{abstract}
\maketitle
\section{Introduction} \label{I}
In the field of quantum optics, the electric-dipole interaction of electromagnetic fields with matter is widely studied, while the magnetic-dipole interaction is often neglected \cite{meystre2007elements,scully1999quantum}. This is because in most cases, the electric-dipole interaction is much stronger than the magnetic-dipole one. 
However, when electromagnetic fields interact with magnetic materials with very high electron spin density, the magnetic-dipole interaction dominates~\cite{papasimakis2016electromagnetic}. Yttrium iron garnet (YIG) crystals, a class of ferrimagnetic materials with low-loss and high-spin-density, have attracted much attention in recent years~\cite{bhoi2019photon,yuan2022quantum,rameshti2022cavity,zhang2016cavity}. In particular, the magnon modes excited in YIG crystals and microwave photons can realize the cavity-magnon polaritons and the vacuum Rabi splitting \cite{huebl2013high,zhang2014strongly}, which induced the creation of cavity-magnon systems that brought quantum optics and magnetism researchers together to develop the integration of quantum physics and spintronic technologies  \cite{bhoi2019photon}. 

The photon-magnon (P-M) coupling induced by the magnetic-dipole interaction in the cavity-magnon system links some of the most exciting concepts in modern physics and has been experimentally implemented  \cite{bhoi2019photon,yuan2022quantum,rameshti2022cavity}. More recently, many interesting quantum effects have been studied based on such hybrid systems, including the magnon-photon (magnon) entanglement \cite{li2018magnon,li2019entangling,zhang2019quantum,mousolou2021magnon,cheng2021tripartite}, magnon chaos \cite{xu2020magnon,liu2019phase}, magnon blocking \cite{yan2020magnon,liu2019magnon}, magnon-induced transparency \cite{ullah2020tunable,kong2019magnon, ZCS}, bistability \cite{wang2018bistability,pan2022bistability,yang2021bistability}, Kerr effect \cite{wang2016magnon,zhang2019theory,kong2019magnon}, to name a few. 
Importantly, these phenomena are closely related to the magnetic-dipole interaction strength. A recent review article compared the cavity-magnon systems of different structures, giving different ranges of values for their P-M coupling strengths~\cite{bhoi2019photon}. 
From a theoretical viewpoint, grasping the P-M coupling requires simultaneous solving of Maxwell’s equations and the Landau–Lifshitz–Gilbert equation \cite{bhoi2019photon,prabhakar2009spin}. Additionally, exploring the P-M coupling is also key to building hybrid cavity-magnon systems for quantum communication technology and realizing potential docking with quantum information science. 
Consequently, accurate knowledge of the P-M coupling strength is an essential task, extremely important both for understanding magnetic-dipole interaction and for technical applications, and determines the depth of exploration in the cavity-magnonics field. 
However, the direct measurement of P-M coupling strength is a huge challenge, costly and even impossible to achieve. A wise choice is to indirectly estimate the P-M coupling strength from experimental data on other readily measurable observables, i.e., by resorting to the quantum estimation theory (QET)~\cite{helstrom1969quantum,liu2020quantum,paris2009quantum}. Particularly, this idea has been applied for the estimating coupling strength both for light-matter interactions (Rabi frequency) \cite{bernad2019optimal,chen2019optimal,burgarth2009coupling,romano2007estimation,xie2023quantum}  and optomechanical systems \cite{qvarfort2021optimal,montenegro2022probing,bernad2018optimal,schneiter2020optimal,sanavio2020fisher,carrasco2022estimation,sala2021quantum}.

In recent years, the double-cavity systems have received increasing attention because the auxiliary cavity can facilitate the enhancement/realization of some quantum effects. Examples include facilitating optomechanical ground-state cooling in the unresolved-sideband regime~\cite{liu2018ground}, enabling phonon detection in the optomechanical weak coupling regime, and constructing $\mathcal{PT}$-symmetric systems such as balanced gain-loss coupled cavities~\cite{yang2017enhanced,tchodimou2017distant}.  
It is worth noting that some researchers have also integrated double-cavity system with YIG sphere recently, indicating that the auxiliary cavity can enhance the cavity-magnon quantum correlation \cite{yang2023asymmetric,chen2021perfect,sohail2023enhanced,hidki2022quantifying}, realize nonreciprocal amplification \cite{zhao2022nonreciprocal} and controllable quantum phase transition \cite{qin2022controllable}. 
Naturally, a reasonable prediction is that the auxiliary cavity might be able to assist the parameter estimation task under some conditions.

With such motivations in mind, in the present work, we exploit the quantum Fisher information (QFI) \cite{fisher1925theory,braunstein1994statistical} and classical Fisher information (CFI) \cite{helstrom1969quantum} to investigate the estimation precision limits and measurement strategy of the P-M coupling strength in a driven-dissipative double-cavity-magnon system, where the primary cavity mode individually coupled to magnon and an auxiliary cavity mode.
We explore the effect of various factors such as temperature, loss rate, driving power and detuning on the estimation precision limit.
Remarkably, in comparison with the auxiliary cavity unassisted case, we find that the estimation precision can be greatly improved by appropriately designing the photon tunneling rate. 
Moreover, selecting appropriate Kerr nonlinear coefficients of the magnon can also reduce estimation errors and facilitate parameter estimation tasks. 
Although the fingerprint of the P-M coupling strength is left in the global state of the system, in practical measurements it is almost impossible to access the entire system. To this end, we also investigated how the information about the P-M coupling is distributed in each subsystem.  
The results show that within most of the given parameter regimes, the primary cavity mode is the optimal subsystem for estimating the P-M coupling strength. Further, we considered performing practically Gaussian measurements on it, exploring how much information about the P-M coupling strength can be experimentally extracted.

This paper is organized as follows. In Sec.~\ref{II}, we introduce the driven-dissipative double-cavity-magnon system used to estimate the P-M coupling strength and its steady state quantum fluctuations are then derived by the quantum master equation. In Sec.~\ref{III}, we briefly review some basic formalism about the parameter estimation of Gaussian states, including QFI calculations for Gaussian states and the form of CFI under several Gaussian measurements. In Sec.~\ref{IV}, we numerically examine the influence of various factors on the estimation error and identify the optimal subsystem for estimating the P-M coupling parameter, and further explore the performance of several Gaussian measurement strategies. The last section concludes this article. In Appendix \ref{AA}, the P-M coupling interaction  between the magnons and microwave photons is derived. The stability conditions of the system are given in Appendix \ref{AB}. Appendix \ref{AD} provides the system's  normal mode picture.

\section{Theoretical model and dynamical analysis} \label{II}
\subsection{The model} \label{IIA}
\begin{figure}[hbt!]
	\centering
	\includegraphics[width=1\linewidth]{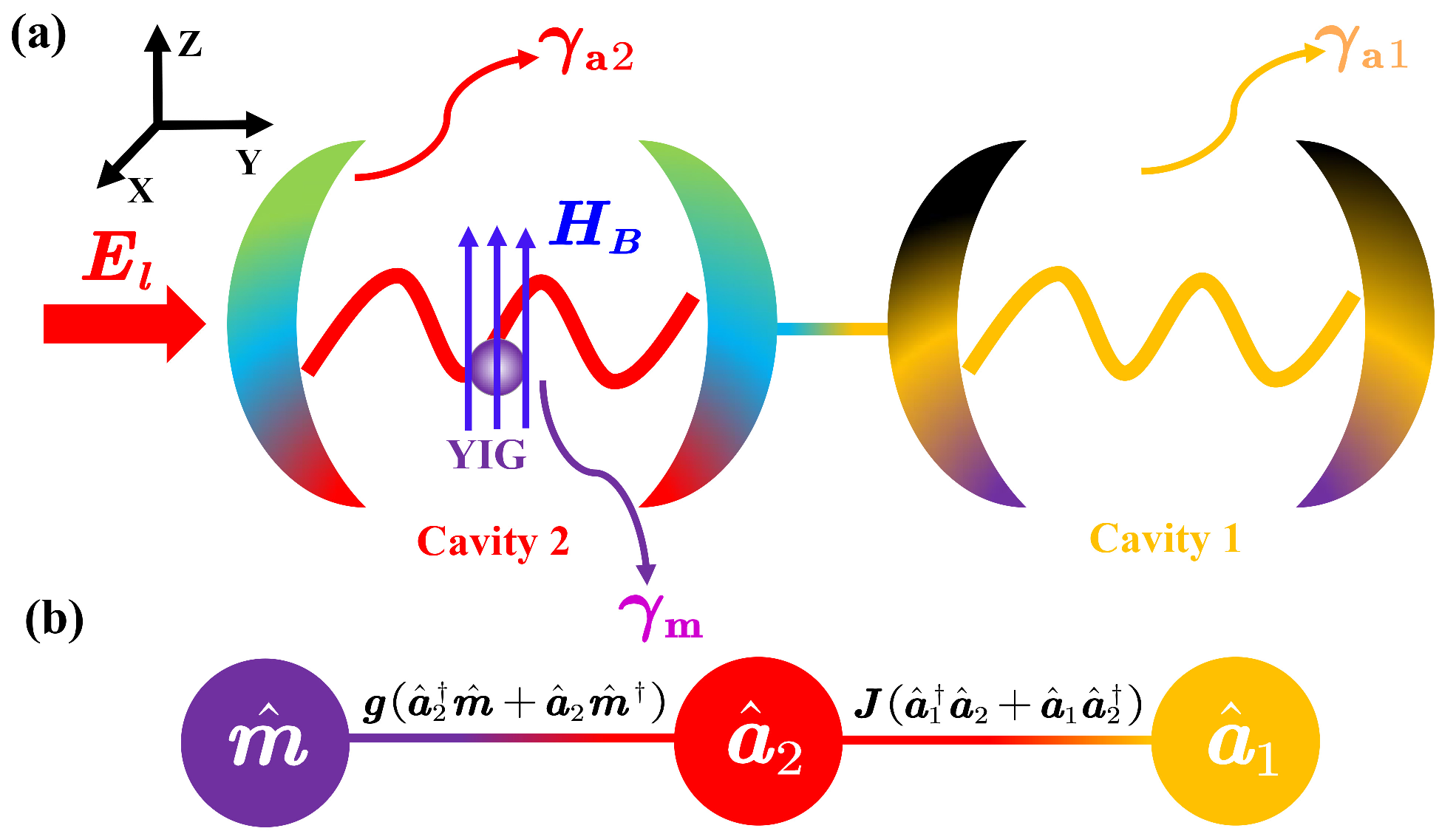}
	\caption{(a) Diagrammatic representation of the driven-dissipative double-cavity-magnon system. The cavity $2$ is driven by a left incident microwave field. Two cavity modes are coupled through photon tunneling. The YIG sphere in the cavity $2$ is magnetized to saturation by an external bias magnetic field $H_{B}$ aligned along the Z direction, which results in the excited magnon modes in the YIG sphere coupled to the cavity modes $2$ via the P-M coupling interaction. (b) The diagram of interactions among
	subsystems in such a hybrid cavity-magnon system.}
	\label{fig1}
\end{figure}

The double-cavity-magnon system we proposed, sketched in Fig.~\ref{fig1}, consists of a highly polished micrometer-scale YIG sphere and two three-dimensional (3D) copper microwave cavities, where the YIG sphere is trapped in the primary cavity $2$ and the cavity $1$ serves as an auxiliary cavity. 
The two cavity modes are coupled to each other by a photon tunneling interaction with a hopping rate $J$. 
In realistic systems, it can be realized through optical backscattering, which depends on the material defects and surface roughness in experimental devices \cite{chen2021synthetic}. Under the action of external static-uniform bias magnetic field $H_{B}$ along the Z-axis, the  YIG sphere will excite many magnons, and the magnon modes exhibit  uniform  spin precession in the YIG sphere \cite{bhoi2019photon,yuan2022quantum,rameshti2022cavity}.  At the same time, the Kerr nonlinear effect of the magnons is also induced owing to the magnetocrystalline anisotropy~\cite{bhoi2019photon,yuan2022quantum,rameshti2022cavity}. In addition, the magnon mode will couple with the cavity mode $2$ via the  beam-splitter-liked P-M coupling with coupling strength $g$.

Here, the P-M coupling rate $g$ is the physical parameter we are interested in, i.e., the parameter to be estimated. We consider a  microwave field along the Y-axis with the power $P_{l}$ and the frequency $\omega _{l}$ to directly drive the cavity $2$. As such, the total Hamiltonian of the hybrid cavity-magnon system can be written as  \cite{yuan2022quantum,rameshti2022cavity} (we set $\hbar =1$ hereafter)
\begin{eqnarray}
\label{Eq1}
\hat{H} &=&\omega _{a_{1}}\hat{a}_{1}^{\dag }\hat{a}_{1}+\omega _{a_{2}}\hat{a}%
_{2}^{\dag }\hat{a}_{2}+\omega _{m}\hat{m}^{\dag }\hat{m}+K\hat{m}^{\dag }%
\hat{m}\hat{m}^{\dag }\hat{m}  \nonumber \\
&&+J(\hat{a}_{1}^{\dag }\hat{a}_{2}+\hat{a}_{1}\hat{%
	a}_{2}^{\dag })+g(\hat{a}_{2}^{\dag }\hat{m}+\hat{a}_{2}\hat{m}^{\dag }) 
\nonumber \\
&&+iE_{l}(\hat{a}_{2}^{\dag }e^{-i\omega _{l}t}-\hat{a}_{2}e^{i\omega
	_{l}t}),
\end{eqnarray}%
where $\hat{a}_{1}^{\dag }$ $(\hat{a}_{2}^{\dag })$ and $\hat{a}_{1}$ $(%
\hat{a}_{2})$ denote the bosonic creation and annihilation operators for the two
cavity modes with frequency $\omega _{a_{1}}$ and $\omega _{a_{2}}$, respectively. $\hat{m}^{\dag }$ $(\hat{m})$ being the creation (annihilation) operator for the magnon mode with frequency $\omega _{m}=\gamma _{e}H_{B}-2\mu _{0}K_{\text{an}}\gamma_{e}^{2}S/V_{m}$M$_{b}^{2}$ \cite{bhoi2019photon,yuan2022quantum,rameshti2022cavity}, here $\gamma _{e}/2\pi =28$ GHz/T is the gyromagnetic ratio for electron (other parameters see Appendix \ref{AA}). The fourth item refers to the magnon Kerr effect with Kerr nonlinear coefficient $K$. The fifth and sixth items refer to the photon-photon hoping and the P-M (magnetic-dipole) interaction, respectively. 
The last term represents the driving to the cavity mode $2$ by the microwave field. 
$E_{l}=\sqrt{\gamma _{a_{2}}P_{l}/\omega _{l}}$ being the amplitude of driving field, in which $\gamma _{a_{2}}=\gamma _{a_{2}}^{0}+\gamma _{a_{2}}^{\text{ex}}$ is the total linewidth of cavity mode $2$.
The linewidth originate from two parts: the intrinsic loss of the cavity mode with rate $\gamma _{a_{2}}^{0}$ (e.g, absorption inside the cavity dielectric) and the photon loss from the open port with rate $\gamma _{a_{2}}^{\text{ex}}$ \cite{aspelmeyer2014cavity}. 
In addition, as in many theoretical or experimental articles~\cite{wang2018bistability,pan2022bistability,yang2021bistability,wang2016magnon,zhang2019theory}, in Eq.~(\ref{Eq1}) we do not include the magnetostrictive effect of YIG sphere because it is very weak. On the other hand, the radiation pressure effect of microwave field on YIG sphere is also ignored. This stems from the fact that the size of the YIG sphere is much smaller than the wavelength of microwave (e.g., the wavelength of microwave with a frequency of 10 GHz is approximately 3 cm which is much larger than the micrometer-scale of YIG sphere). 
 
Defining the vector of operators $\mathbf{\hat{r}}:=[\hat{a}_{1},\hat{a}_{1}^{\dag },\hat{a}_{2},\hat{a}_{2}^{\dag },\hat{m},\hat{m}^{\dag }]^{\text{T}}$, the commutation relations between the operators satisfy $[\mathbf{\hat{r}}_{j},\mathbf{\hat{r}}_{k}]=\Xi _{jk}$. Here $\Xi $ is a $%
6\times 6$-dimensional symplectic matrix, which enforces the uncertainty relations in the phase-space  \cite{aspelmeyer2014cavity,serafini2023quantum}, and reads as 
\begin{equation}
\Xi :=\overset{3}{\underset{k=1}{\oplus }}\Lambda ,\Lambda :=\left[ 
\begin{array}{cc}
0 & 1 \\ 
-1 & 0%
\end{array}%
\right] .
\end{equation}%
It is straightforward to verify that $\Xi ^{\text{T}}=\Xi ^{-1}=-\Xi $. 

To eliminate the time factor in Eq.~(\ref{Eq1}), the rotating frame with respect to the frequency of the driving microwave $\omega _{l}$ is applied, the Hamiltonian can be rewritten as
\begin{eqnarray}
\hat{H}_{\text{r}}&=&\Delta _{a_{1}}\hat{a}_{1}^{\dag }\hat{a}_{1}+\Delta _{a_{2}}\hat{a}_{2}^{\dag }%
\hat{a}_{2}+\Delta _{m}\hat{m}^{\dag }\hat{m}+K\hat{m}^{\dag }\hat{m}\hat{m}%
^{\dag }\hat{m}  \notag \\
&&+J(\hat{a}_{1}^{\dag }\hat{a}_{2}+\hat{a}_{1}\hat{%
	a}_{2}^{\dag })+g(\hat{a}_{2}^{\dag }\hat{m}+\hat{a}_{2}\hat{m}^{\dag }) 
\notag \\
&&+iE_{l}(\hat{a}_{2}^{\dag }-\hat{a}_{2}),
\end{eqnarray}%
where $\Delta _{i}=\omega _{i}-\omega _{l}$ $(i=a_{1},a_{2},m)$ denotes the detuning
of driving microwave from the mode $i$, in which $\Delta _{i}>0$ and  $\Delta _{i}<0$ refer to red detuning and blue detuning, respectively.

\subsection{Dynamical analysis} \label{IIB}
In addition to the driving term, also considering Gaussian environment, so the cavity modes losses and the magnon  damping will also be included in dynamic evolution. Our model is thus essentially a driven-dissipative double-cavity-magnon system. To observe dynamical behavior of the system,
one can exploit the Lindblad master equation \cite{breuer2002theory}, which offers a powerful
framework for simulating open systems, i.e.,
\begin{equation}
\label{Eq4}
\frac{d\hat{\rho}(t)}{dt}=-i[\hat{H}_{\text{r}},\hat{\rho}(t)]+\sum%
\limits_{i,j=1}^{6}\dfrac{\Gamma _{ij}}{2}[2\mathbf{\hat{r}}_{i}\hat{\rho (t)%
}\mathbf{\hat{r}}_{j}-\{\mathbf{\hat{r}}_{j}\mathbf{\hat{r}}_{i},\hat{\rho}%
(t)\}],
\end{equation}%
where $\hat{\rho}(t)$ denotes the density matrix of the system; $\{\mathbf{\hat{r}}_{i},\mathbf{\hat{r}}_{j}\}\in \lbrack \{\hat{a}_{1},\hat{%
	a}_{1}^{\dag }\},\{\hat{a}_{2},\hat{a}_{2}^{\dag }\},\{\hat{m},\hat{m}^{\dag
}\}]$  and  $\{\mathbf{\hat{r}}_{i},\mathbf{\hat{r}}_{j}\}\in \lbrack \{\hat{a}%
_{1}^{\dag },\hat{a}_{1}\},\{\hat{a}_{2}^{\dag },\hat{a}_{2}\},\{\hat{m}%
^{\dag },\hat{m}\}]$ represent the particle losses and the phase insensitive linear amplification processes, respectively \cite{breuer2002theory}; $\Gamma =\Gamma _{a_{1}}\oplus \Gamma _{a_{2}}\oplus \Gamma _{m}$ being the damping
matrix, in which
\begin{equation}
\Gamma _{k=a_1, a_2, m}=\left[ 
\begin{array}{cc}
0 & \gamma _{k}[n(\omega _{k})+1] \\ 
\gamma _{k}n(\omega _{k}) & 0%
\end{array}%
\right] ,
\end{equation}%
where $\gamma_{k}$ is the decay rate of mode $k$, and the bose number $n(\omega _{k})=[\exp (\omega_{k}/k_{B}T)-1]^{-1}$ is the mean occupancy of the mode $k$ wherein $T$ is the environment temperature and $k_{B}$ the Boltzmann constant  
\cite{aspelmeyer2014cavity,serafini2023quantum}.
Notice that for optical frequencies (about 10 THz $\sim 10^{4}$ THz) $n(\omega _{a_{1},a_2})$ can be ignored at
room temperature, however, due to the cavity modes under consideration are at microwave frequencies (The experimental frequency range of cavity modes in cavity-magnon system is on the order of GHz, due to the lack of efficient THz radiation sources and corresponding detection electronics \cite{bhoi2019photon}), $n(\omega _{a_{1},a_2})$ can be
comparable to mean thermal magnon number $n(\omega _{m})$ \cite{wilson2008cavity}.  

Suppose the driving microwave field is relatively strong,
its main effect is to displace the steady state of all modes. As a result, each operator can be safely considered as a small quantum fluctuation above a steady state value~\cite{aspelmeyer2014cavity,serafini2023quantum}, i.e.,  
\begin{equation}
\label{Eq6}
\mathbf{\hat{r}\rightarrow \langle \hat{r}_{0}\rangle +}\delta \mathbf{\hat{r}},
\end{equation}%
where $\mathbf{\langle \hat{r}_{0}\rangle =}$ Tr$[\mathbf{\hat{r}}\hat{\rho}(\infty )]$ denoting the vector of steady state averages and $\delta \mathbf{\hat{r}}$ is the quantum fluctuation vector around the steady state value. 
According to Eq.~(\ref{Eq4}), the equation of motion for the average value of an arbitrary operator $\hat{\mathcal{O}} $ is given by \cite{breuer2002theory,puri2001mathematical}, 
\begin{equation}
\frac{d\mathbf{\langle }\hat{\mathcal{O}}\mathbf{\rangle }}{dt}=-i\langle \lbrack \hat{%
	\mathcal{O}},\hat{H}_{\text{r}}]\rangle +\sum\limits_{i,j=1}^{6}\dfrac{\Gamma _{ij}}{2}%
\langle \lbrack \mathbf{\hat{r}}_{j},\hat{\mathcal{O}}]\mathbf{\hat{r}}_{i}-\mathbf{%
	\hat{r}}_{j}[\mathbf{\hat{r}}_{i},\hat{\mathcal{O}}]\rangle ,
\end{equation}%
where the cyclic property of the matrix's trace is used. 
Selecting $\hat{\mathcal{O}}\in \mathbf{%
	\hat{r}}_{i}$ and setting $d\mathbf{\langle }\hat{\mathcal{O}}\mathbf{\rangle /}dt\equiv 0$
for steady state ($t\rightarrow \infty $), one can get
\begin{eqnarray}
\mathbf{\langle }\hat{a}_{1}\mathbf{\rangle } &\mathbf{=}&\frac{%
	-iJ\mathbf{\langle }\hat{a}_{2}\mathbf{\rangle }}{i\Delta
	_{a_{1}}+\gamma _{a_{1}}}, \\
\mathbf{\langle }\hat{a}_{2}\mathbf{\rangle } &\mathbf{=}&\frac{(E_{l}-ig%
	\mathbf{\langle }\hat{m}\mathbf{\rangle })(i\Delta _{a_{1}}+\gamma _{a_{1}})%
}{(i\Delta _{a_{1}}+\gamma _{a_{1}})(i\Delta _{a_{2}}+\gamma _{a_{2}})+J^{2}}%
, \\
\mathbf{\langle }\hat{m}\mathbf{\rangle } &\mathbf{=}&\frac{-igE_{l}(i\Delta
	_{a_{1}}+\gamma _{a_{1}})}{\mathcal{Q}[(i\Delta _{a_{1}}+\gamma _{a_{1}})(i\Delta
	_{a_{2}}+\gamma _{a_{2}})+J^{2}]}, \label{Eq10}
\end{eqnarray}%
with
\begin{eqnarray}
\mathcal{Q} &=&i(\Delta _{m}+2K|\mathbf{\langle }\hat{m}\mathbf{\rangle }%
|^{2}+K)+\gamma _{m}+  \notag \\
&&\frac{g^{2}(i\Delta _{a_{1}}+\gamma _{a_{1}})}{(i\Delta _{a_{1}}+\gamma
	_{a_{1}})(i\Delta _{a_{2}}+\gamma _{a_{2}})+J^{2}}.
\end{eqnarray}
Correspondingly, the strong driving assumption of microwave field is equivalent to the mean photon number of cavity mode $2$ is large, i.e., $| \langle \hat{a}_{2}\rangle |^{2}\gg 1$. 
Note that the system may exhibit multiple steady state solutions, e.g., Eq.~(\ref{Eq10}) is a unary cubic equation
about the mean magnon number $|\mathbf{\langle }\hat{m}\mathbf{\rangle }|^{2}$. In current work, we only focus on parameter regimes in which the system does not exhibit multistability, namely all modes have unique steady state solutions. This is equivalent to setting a limit on the driving strength.

Further, applying linearization approximation Eq.~(\ref{Eq6}) to Eq.~(\ref{Eq4}), one can get the bilinear quantum master equation,
\begin{equation}
\label{Eq12}
\frac{d\hat{\rho}(t)}{dt}=-i[\hat{H}_{\text{eff}},\hat{\rho}(t)]+\sum%
\limits_{i,j=1}^{6}\dfrac{\Gamma _{ij}}{2}[2\delta \mathbf{\hat{r}}_{i}\hat{%
	\rho (t)\delta }\mathbf{\hat{r}}_{j}-\{\delta \mathbf{\hat{r}}_{j}\delta \mathbf{\hat{r}}_{i},\hat{\rho}(t)\}]
\end{equation}
with linearized effective Hamiltonian
\begin{eqnarray}
\label{Eq13A}
\hat{H}_{\text{eff}} &=&\Delta _{1}\delta \hat{a}_{1}^{\dag }\delta \hat{a}%
_{1}+\Delta _{2}\delta \hat{a}_{2}^{\dag }\delta \hat{a}_{2}+\Delta
_{\text{eff}}\delta \hat{m}^{\dag }\delta \hat{m}  \notag \\
&&+K\left[\mathbf{\langle }\hat{m}\mathbf{\rangle }^{2}\delta \hat{m}^{\dag
}\delta \hat{m}^{\dag }+\mathbf{\langle }\hat{m}\mathbf{\rangle }^{\ast
	^{2}}\delta \hat{m}\delta \hat{m}\right]\ + \\
&&J(\delta \hat{a}_{1}^{\dag }\delta \hat{a}_{2}+\delta \hat{a}_{1}\delta 
\hat{a}_{2}^{\dag })+g(\delta \hat{a}_{2}^{\dag }\delta \hat{m}+\delta \hat{a%
}_{2}\delta \hat{m}^{\dag }),  \notag
\end{eqnarray}%
where only the quadratic order terms of fluctuations are retained and $\Delta _{\text{eff}}=\Delta
_{m}+4K|\mathbf{\langle }\hat{m}\mathbf{\rangle }|^{2}$ is the effective detuning of magnon in the presence of Kerr effect. 

In particular, the current work strongly relies on the framework of Gaussian
state, whence it convenient to model our system by dimensionless quadrature
operators in the phase-space. Defining Hermitian fluctuation quadrature operators  \cite{aspelmeyer2014cavity,serafini2023quantum}
\begin{eqnarray}
\delta \hat{Q}_{o} &:=&(\delta \hat{o}+\delta \hat{o}^{\dag })/\sqrt{2}, \\
\delta \hat{P}_{o} &:=&(\delta \hat{o}-\delta \hat{o}^{\dag })/\sqrt{2}i,
\end{eqnarray}%
where $o=a_{1},a_{2}$ and $m$. The Lyapunov equation for the steady state
covariance matrix $\mathcal{V}$ can be obtained through Eq. (\ref{Eq12}), i.e.,
\begin{equation}
\label{EQ16}
\mathfrak{A}\mathcal{V}+\mathcal{V}\mathfrak{A}^{\text{T}}=-\mathfrak{D},
\end{equation}%
with
\begin{equation}
\mathfrak{A} =\left[ 
\begin{array}{cccccc}
-\gamma _{a_{1}} & \Delta _{a_{1}} & 0 & J & 0
& 0 \\ 
-\Delta _{a_{1}} & -\gamma _{a_{1}} & -J & 0 & 
0 & 0 \\ 
0 & J & -\gamma _{a_{2}} & 
\Delta _{a_{2}} & 0 & g \\ 
-J & 0 & -\Delta _{a_{2}} & 
-\gamma _{a_{2}} & -g & 0 \\ 
0 & 0 & 0 & g & \Re _{+} & \Im _{+} \\ 
0 & 0 & -g & 0 & \Im _{-} & \Re _{-}%
\end{array}%
\right] , 
\end{equation}
and
\begin{eqnarray}
\mathfrak{D} &=&\text{diag}\{[2n(\omega _{a_{1}})+1]\gamma _{a_{1}},[2n(\omega
_{a_{1}})+1]\gamma _{a_{1}},  \notag \\
&&[2n(\omega _{a_{2}})+1]\gamma _{a_{2}},[2n(\omega _{a_{2}})+1]\gamma
_{a_{2}},  \notag \\
&&[2n(\omega _{m})+1]\gamma _{m},[2n(\omega _{m})+1]\gamma _{m}\},
\end{eqnarray}
where $\mathfrak{A}$ and $\mathfrak{D}$ are the drift and  diffusion matrices, respectively; $\Re _{\pm }=-\gamma _{m}\pm 2$Im$(\mathbf{\langle }%
\hat{m}\mathbf{\rangle }^{2})K$ and $\Im _{\pm }=\pm (\Delta _{\text{eff}}+K)-2$Re$(%
\mathbf{\langle }\hat{m}\mathbf{\rangle }^{2})K$. Here, we point out that Eq.~(\ref{EQ16}) is a linear equation and can be solved analytically. However, the explicit solution is cumbersome, so we utilize numerical solutions below. 

The covariance matrix $\mathcal{V}$'s $ij$-element are given by $\mathcal{V}_{ij}:=\langle \mathbf{%
	\delta\hat{R}}_{i}(\infty )\mathbf{\delta\hat{R}}_{j}(\infty )+\mathbf{\delta\hat{R}}%
_{j}(\infty )\delta\mathbf{\hat{R}}_{i}(\infty )\rangle /2$, where fluctuation
quadrature operators vector $\delta\mathbf{\hat{R}}:=[\delta \hat{Q}_{a_{1}},\delta 
\hat{P}_{a_{1}},\delta \hat{Q}_{a_{2}},\delta \hat{P}_{a_{2}},\delta \hat{Q}%
_{m},\delta \hat{P}_{m}]^{\text{T}}$ is defined and satisfying commutation relations $[\mathbf{\delta\hat{R}}_{j},\mathbf{\delta\hat{R}}_{k}]=i\Xi _{jk}$. This indicates that the steady second moment of the system is encoded on the covariance matrix $\mathcal{V}$, formally
\begin{equation}
\label{eq19g}
\mathcal{V}:=\left[ 
\begin{array}{ccc}
L_{a_{1}} & C_{a_{1},a_{2}} & C_{a_{1},m} \\ 
C_{a_{2},a_{1}}^{\text{T}} & L_{a_{2}} & C_{a_{2},m} \\ 
C_{m,a_{1}}^{\text{T}} & C_{m,a_{2}}^{\text{T}} & L_{m}%
\end{array}%
\right] ,
\end{equation}%
where $L_{i}$ and $C_{i,j}$ being a $2\times 2$ subblock matrices of $\mathcal{V}$ $%
(i,j=a_{1},a_{2},m)$, they represent the local properties of mode $i$ and the quantum correlation between modes $i$ and $j$, respectively. 
The covariance matrix $\mathcal{V}$ is a real symmetric matrix and positive semidefinite, satisfying the Robertson-Schrödinger uncertainty principle $\mathcal{V}+i\Xi/2 \geq 0$  \cite{aspelmeyer2014cavity,serafini2023quantum}.  
This uncertainty relationship is a necessary and sufficient condition for $\mathcal{V}$ to be a stable covariance matrix for Gaussian states. According to the Routh-Hurwitz criterion, the system is far away from instabilities and multistabilities if and only if the real part of all eigenvalues of drift matrix $\mathfrak{A}$  (Lyapunov exponents) are negative \cite{dejesus1987routh}. Otherwise,
the steady state of the system will not converge to a fixed point and chaos or limit cycle phenomena may occur. 
In the present work, we have selected proper parameters to satisfy the specific stability condition shown in Appendix \ref{AB}. 

\section{Parameter estimation theory for quantum Gaussian states} \label{III}
\subsection{Quantum Fisher information for Gaussian states} \label{IIIA}
In particular, Gaussian initial state still remains in Gaussian after it
follows the evolution of bilinear master equation, which is called a
Gaussian transformation \cite{aspelmeyer2014cavity,serafini2023quantum}. The steady state of the proposed system
must be therefore Gaussian owing to the linearized effective Hamiltonian. Importantly, the characteristic function $\mathbf{\chi }_{G}(\mathbf{\widetilde{R}})$ of a generic Gaussian state is completely characterized by two important statistical quantities,
that is first moment (displacement vector) $\mathbf{\langle \hat{R}}_{0}\mathbf{\rangle }$ and
covariance matrix $\mathcal{V}$, where $\mathbf{\langle \hat{R}}_{0}\mathbf{\rangle }%
:=[\mathbf{\langle }\hat{Q}_{a_{1}}\mathbf{\rangle },\mathbf{\langle }\hat{P}%
_{a_{1}}\mathbf{\rangle },\mathbf{\langle }\hat{Q}_{a_{2}}\mathbf{\rangle },%
\mathbf{\langle }\hat{P}_{a_{2}}\mathbf{\rangle },\mathbf{\langle }\hat{Q}%
_{m}\mathbf{\rangle },\mathbf{\langle }\hat{P}_{m}\mathbf{\rangle }]^{\text{T%
}}$ being the vector of steady state quadrature average. Specifically,  $\mathbf{\chi }_{G}(\mathbf{\widetilde{R}})$ is written as 
\begin{equation}
\chi _{G}(\mathbf{\widetilde{R}}):=\text{exp}\left[ -\frac{1}{2}\mathbf{\widetilde{R}}^{\text{T}%
}\mathcal{V}\mathbf{\widetilde{R}+}i\mathbf{\widetilde{R}}^{\text{T}}\langle \mathbf{\hat{R}}%
_{0}\rangle \right], 
\end{equation}
where real variable vector $\mathbf{\widetilde{R}}= \Xi \mathbf{R}$ is defined, in which $\mathbf{R}%
:=[Q_{a_{1}},P_{a_{1}},Q_{a_{2}},P_{a_{2}},Q_{m},P_{m}]^{\text{T}}$. The range of value for each element in $\mathbf{R}$ is $\left[ -\infty ,+\infty \right]$. The all relevant
parameters information of the cavity-magnon system  is thus fully included in $%
\mathbf{\langle \hat{R}}_{0}\mathbf{\rangle }$ and $\mathcal{V}$. It is natural to think of using the Gaussian state’s moments to express QFI. In this sense, the task of estimating the parameter information of an \emph{infinite-dimensional} system is transformed into dealing with the problem of \emph{finite-dimensional} moments.

The definition of the QFI  for the P-M coupling parameter $g$ is 
\begin{equation}
\mathcal{F}_{g}:=\text{Tr}[\hat{\rho}_{g}\hat{L}_{g}^{2}],
\end{equation}
where $\hat{\rho}_{g}$ being the $g$-dependent density matrix of the system; $\hat{L}_{g}$  is the symmetric logarithmic derivative, defined in a way that $2\partial _{g}\hat{\rho}_{g}=\hat{\rho}_{g}\hat{L}_{g}+\hat{L}_{g}\hat{\rho}_{g}$. Based on the definitions of $\mathbf{\chi }_{G}(\mathbf{\widetilde{R}})$, $\mathcal{F}_{g}$, and $\hat{L}_{g}$, after tedious algebra one find that $\mathcal{F}_{g}$ can be rewritten as \cite{bakmou2020multiparameter,pinel2012ultimate,monras2013phase,vsafranek2018estimation}
\begin{equation}
\label{Eq20}
\mathcal{F}_{g}=2\text{vec}[\partial _{g}\mathcal{V}]^{\dag } \mathfrak{M}^{-1}\text{vec}%
[\partial _{g}\mathcal{V}]+\partial _{g}\mathbf{\langle \mathbf{\hat{R}}_{0}\rangle }%
^{\text{T}}\mathcal{V}^{-1}\partial _{g}\mathbf{\langle \mathbf{\hat{R}}_{0}\rangle ,}
\end{equation}%
where we have defined $\partial _{g}\equiv \partial /\partial _{g}$; $\mathfrak{M}=(4\mathcal{V}^{\dag }\otimes \mathcal{V}+\Xi \otimes \Xi )$; vec$[\mathfrak{G}]$ denotes the
vectorization of a matrix $\mathfrak{G}$ ($n$-dimension), which is defined as vec$[\mathfrak{G}]:=\left[ \mathfrak{G}\left( :,1\right) ^{\text{T}},\mathfrak{G}\left( :,2\right) ^{\text{T}%
},\cdots ,\mathfrak{G}\left( :,n\right) ^{\text{T}}\right] ^{\text{T}}$, and $\mathfrak{G}\left( :,n\right)$ being the $n$-th column of $\mathfrak{G}$. Interestingly, one can see that its first term is the contribution owing to the dependence of the second moment on the estimated
parameter, while the second term is the contribution originating from the dependence
of the first moment on the estimated parameter. Correspondingly, the ultimate
precision limit of estimating $g$ is quantified by the quantum Cram\'er-Rao bound (QCRB)  inequality, i.e., \cite{helstrom1969quantum,liu2020quantum,paris2009quantum}
\begin{equation}
\text{Var}\left( \hat{g}\right) \geq \frac{1}{\mathcal{N}\mathcal{F}_{g}},
\end{equation}%
where Var$(\hat{g})=\langle (\hat{g}-g)^{2}\rangle $ being the mean-square
error of unbiased estimator (a statistic that satisfies $\langle \hat{g}\rangle =g
$) for the parameter $g$; $\mathcal{N}$ is the number of independent repetition of the 
estimation protocol or equivalently, the number of independent probes. This indicates that the larger $\mathcal{N}$ and $\mathcal{F}_{g}$, the theoretically higher precision can be achieved for estimating the P-M coupling parameter.

Equation (\ref{Eq20}) presents the QFI obtained by extracting the P-M coupling's information
based on the global state of the system. Its extraction usually requires complex joint measurements on subsystems (cavity modes $1$ and $2$, magnon mode). In practical experiments, it is more feasible to measure only one of the subsystems. 
The interaction between subsystems leads to the transfer of interested parameter information between them. As a consequence, the fingerprint of the P-M coupling strength $g$ is left in each subsystem. 
Thus, two questions naturally arise: (1) what is the precision limit for estimating $g$
based on each subsystem; (2) which is the optimal subsystem for estimating the P-M coupling strength (defined as the subsystem that contains the most information about $g$ under the same conditions). 
To this end, we introduce the QFI for each subsystem
\begin{equation}
\mathcal{F}_{g}^{i}=2\text{vec}[\partial _{g}L_{i}]^{\dag }M_{i}^{-1}\text{vec}%
[\partial _{g}L_{i}]+\partial _{g}\mathbf{\langle }\hat{d}_{i}\mathbf{%
	\rangle }^{\text{T}}L_{i}^{-1}\partial _{g}\mathbf{\langle }\hat{d}_{i}%
\mathbf{\rangle ,}
\end{equation}%
with 
\begin{equation}
M_{i}=(4L_{i}^{\dag }\otimes L_{i}+\Lambda \otimes \Lambda ),\mathbf{\langle }%
\hat{d}_{i}\mathbf{\rangle }\mathbf{=[\mathbf{\langle }}\hat{Q}_{i}\mathbf{%
	\mathbf{\rangle ,\langle }}\hat{P}_{i}\mathbf{\mathbf{\rangle }]^\text{T},}
\end{equation}%
where $i=a_{1}$, $a_{2}$ and $m$; $\mathcal{F}_{g}^{i}$ and $\mathbf{\langle }\hat{d}_{i}\mathbf{\rangle }$  correspond to the QFI and  displacement vector for mode $i$, respectively. As a reminder, $L_{i}$ and $\langle \hat{d}_{i}\rangle$ can be obtained through $\mathcal{V}$ and $\mathbf{\langle \hat{R}}_{0}\mathbf{\rangle }$, respectively. 
Note that the QFI of the global system is usually not equal to the sum of the QFIs of all subsystems.

\subsection{ Classical Fisher information for Gaussian states} \label{IIIB}
For a parameter-dependent quantum state (currently, it refers to the steady state of the double-cavity-magnon system), the QFI sets the theoretical lower bound on the estimation error. However, the actual achievable estimation precision should be evaluated by the classical Fisher information (CFI).
In order to obtain CFI, the positive-operator valued measure (POVM) $\{\hat{\Pi}
_{y}\}\in \widetilde{\Omega}$ (all POVMs) is usually carried out on the parameter-dependent quantum state, where $\{\hat{\Pi}
_{y}\}$ satisfies the completeness condition $\sum_{y}\hat{\Pi} _{y}^{\dag }\hat{\Pi}
_{y}=\mathbb{1}$ and positive semidefinite $\hat{\Pi}
_{y} \geq 0$ \cite{degen2017quantum}.  After performing the measurement, we can obtain a conditional
probability distribution by the Born rule, i.e.,
\begin{equation}
\{P(y|g)=\text{Tr}[\hat{\rho}_{g}\hat{\Pi} _{y}]\}, 
\end{equation}%
where $P(y|g)$ being the conditional probability for the measurement result $y$.  
We then can infer the value of the parameter of interest based on probability distribution of parameter dependence. The corresponding CFI reads as \cite{helstrom1969quantum,liu2020quantum,paris2009quantum}
\begin{equation}
\label{Eq24}
F_{g}:=\int \frac{1}{P(y|g)}\left[ \frac{\partial P(y|g)}{\partial g}%
\right] ^{2}dy.
\end{equation}%
According to the definition of QFI, so $\mathcal{F}_{g}:=$ $\underset{\tilde{\Omega}}{\text{Max}}%
\{F_{g}\}$ holds already optimized over the class of
all possibility POVMs. Hence one have that
\begin{equation}
\text{Var}\left( \hat{g}\right) \geq \frac{1}{\mathcal{N}F_{g}}\geq \frac{1}{\mathcal{N}\mathcal{F}_{g}}.
\end{equation}%
The equal sign in front corresponds to classical CRB, in principle, which can be saturated through optimal estimators, e.g., the maximum likelihood estimator.  
The second equal sign can be saturated by the optimal measurement strategy. We point out that the measurements mentioned in this paper are all ideal, namely the measurement efficiency is $1$.

For single-parameter estimation scenarios, the optimal
measurement setup can be constructed by the eigenbasis of the symmetric
logarithmic derivative $\hat{L}_{g}$. Unfortunately,  however, in real scenarios the optimal measurement setup may be experimentally very demanding and even effectively out of reach. This implies that the
ultimate estimation precision set by QFI may not be truly achievable. It is therefore advisable to investigate experimentally realizable measurement strategies, here we focus on realistic Gaussian measurements.

For the continuous-variable systems, the well-known
detection schemes are Homodyne and Heterodyne detections \cite{meystre2007elements,scully1999quantum}, both belonging to Gaussian measurement, with a prominent status in signal processing. 
When performing Gaussian measurements on a Gaussian state, the output result satisfies a Gaussian distribution. Generally speaking, in the Gaussian measurement process, we can read the parameters of interest through the statistical moments of the quadrature operators. 
Then, based on the error propagation formula, the corresponding estimation precision can be reflected.  In the current work, we will explore which scheme in Homodyne and Heterodyne detections can achieve higher precision for estimating the P-M coupling parameter, and benchmark their performance by comparing them with the ultimate precision limit of the optimal measurement settings.  
Notice that, in practice, it is usually not feasible to perform global measurements on multimode Gaussian states, as this involves joint measurements of all modes, which poses an enormous challenge for experiments. As such, in the subsequent discussion, we focus on the single-mode Gaussian measurements that only partially access one of the subsystems.

\subsubsection{Homodyne detection} \label{IIIB1}
For a Homodyne detection scheme, the
signal light is mixed with a probe (strong local oscillator) at the same frequency. Formally, the measurement projection operators of Homodyne detection are $\{|\hat{X}\rangle
\langle \hat{X}|\}$ or $\{|\hat{Y}\rangle \langle \hat{Y}|\}$, where $|\hat{X}\rangle $ $(|\hat{Y}\rangle )
$ denotes the eigenvector of quadrature operator $\hat{X}$ $(\hat{Y})$ \cite{meystre2007elements,scully1999quantum,serafini2023quantum,oh2019optimal}. At the present, $%
\hat{X}\in \{\hat{Q}_{a_{1}},\hat{Q}_{a_{2}},\hat{Q}_{m}\}$ and $\hat{Y}\in \{\hat{P}%
_{a_{1}},\hat{P}_{a_{2}},\hat{P}_{m}\}$, where the quadrature operators of the cavity modes can be directly measured, while measuring that of the magnon can
be completed by introducing auxiliary lights. For example by coupling the YIG sphere to an additional microwave cavity driven by a weak field. This leads to the state-swap interaction which maps the magnon state onto the cavity output field. 

The measurement outcomes $X$ and $Y$ satisfy the Gaussian probability distribution, and are given by
\begin{eqnarray}
\label{Eq26}
P(X) &:&=\int \mathcal{W}(X,Y)dY, \\
P(Y) &:&=\int \mathcal{W}(X,Y)dX, \label{Eq27}
\end{eqnarray}%
where $\mathcal{W}(X, Y)$ denotes the Wigner function of the Gaussian state, which depends on the first and second moments of the system. 
Based on Eqs.~(\ref{Eq26})-(\ref{Eq27}), we can obtain the Gaussian distribution associated with the measurement of quadratures of the double-cavity-magnon system \cite{serafini2023quantum},
\begin{equation}
\label{Eq28}
P(\mathbf{R}_{k},g)=\frac{\text{exp}[-(\mathbf{R}_{k}-\langle  \mathbf{\hat{R}}_{0}\rangle_{k} )^{2}/2\mathcal{V}_{kk}]}{\sqrt{2\pi \mathcal{V}_{kk}}}, 
\end{equation}
where $\mathbf{R}_{k}$ and $\langle \mathbf{\hat{R}}_{0}\rangle_{k} $ are real variable and the steady state value of the selected quadrature operator ($ k=1\sim 6,\text{corresponds sequentially to }\hat{Q}_{a_{1}},\hat{P}%
_{a_{1}},\hat{Q}_{a_{2}},\hat{P}_{a_{2}},\hat{Q}_{m},\hat{P}_{m} $), respectively; $\mathcal{V}_{kk}$ being the diagonal element of the  covariance
matrix $\mathcal{V}$. For instance, when 
$k=4$, we have that  $\mathbf{R}_{4}=P_{a_{2}}$, $\langle \mathbf{\hat{R}}%
_{0}\rangle_{4} =\langle \hat{P}_{a_{2}}\rangle $, and $\mathcal{V}_{44}=\langle
\lbrack \delta \hat{P}_{a_{2}}\left( \infty \right) ]^{2}\rangle $. 
By plugging Eq.~(\ref{Eq28}) into Eq.~(\ref{Eq24}), one can obtain the CFI about $g$ under the Homodyne detection, i.e.,  \cite{monras2013phase}
\begin{equation}
\label{Eq31}
F_{g,\text{Ho}}^{k}=\frac{1}{2\mathcal{V}_{kk}^{2}}\left[ 2\mathcal{V}_{kk}(\partial
_{g}\langle \mathbf{\hat{R}}_{0}\rangle _{k})^{2}+\left( \partial _{g}%
\mathcal{V}_{kk}\right) ^{2}\right],
\end{equation}%
where the superscript  “$k$” refers to the quadrature operator of measurement; the subscript “Ho”  marks the Homodyne detection.

\subsubsection{Heterodyne detection} \label{IIIB2}
Heterodyne detection is also an important Gaussian measurement strategy, where the measured field is mixed with a probe field at a different frequency. Formally, the measurement projection operators of Heterodyne detection constructed from the coherent state, namely $\{|\alpha \rangle \langle \alpha |/\pi \}$ \cite{meystre2007elements,scully1999quantum,serafini2023quantum,oh2019optimal}. 
In general, the Heterodyne detector combines the measured mode with an ancillary vacuum mode into a $50:50$ beam-splitter and then measures the
quadratures $\hat{X}$ and $\hat{Y}$ of the outcome mode. Performing an ideal Heterodyne detection on the subsystems of the cavity-magnon system, one can obtain the probability distribution as \cite{serafini2023quantum}
\begin{equation}
\label{Eq29}
P(d_{i},g)=\frac{\text{exp}\left[ -\frac{1}{2}(d_{i}-\langle \hat{d}_{i}\rangle )^{%
		\text{T}}\aleph_{i} ^{-1}(d_{i}-\langle \hat{d}_{i}\rangle )\right] }{ %
	\sqrt{2\pi\text{det}(\aleph_{i} )}},
\end{equation}%
where $\aleph_{i}=L_{i} +\mathbb{1}_{2}$; $\langle \hat{d}_{i}\rangle $ and $L_{i}$ are the first moment and the covariance matrix of the selected mode, respectively. 
Particularly, $\mathbb{1}_{2}$ is a $2\times 2$ identity matrix, which represents the added noise in $L_{i}$ stems from the simultaneous detection of the conjugated quadrature operators $\hat{X}$ and $\hat{Y}$. 
For example, if Heterodyne detection is considered for the cavity mode $2$, they take the form, respectively, of $\langle \hat{d}_{i}\rangle  =[\mathbf{\langle }\hat{Q}_{a_{2}}\mathbf{%
	\rangle },\mathbf{\langle }\hat{P}_{a_{2}}\mathbf{\rangle }]^{\text{T}}$ and $%
L_{i} =L_{a_{2}}$. Substituting Eq.~(\ref{Eq29}) into
Eq.~(\ref{Eq24}), one can obtain the CFI \cite{monras2013phase},
\begin{equation}
\label{Eq33}
F_{g, \text{He}}^{i}=\frac{1}{2}\text{Tr}\left[ (\aleph ^{-1}\partial _{g}\aleph
)^{2}\right] +\partial _{g}\langle \hat{d}_{i}\rangle ^{\text{T}}\aleph
^{-1}\partial _{g}\langle\hat{d}_{i}\rangle. 
\end{equation}
where “Tr” denotes performing a trace on a matrix; the superscript “$i$” refers to the measured mode; the subscript “He” marks the Heterodyne measurement.

\subsubsection{Optimal Gaussian measurement} \label{IIIB3}
More generally, the CFIs obtained by performing arbitrary Gaussian measurements on
Gaussian states can be uniformly expressed as \cite{cenni2022thermometry}
\begin{eqnarray}
\label{Eq34}
F_{g}^{GM}(\mathbf{d},\bm{\sigma };\bm{\sigma }_{GS}^{M})
&=&\partial _{g}\mathbf{d}^{\text{T}}(\bm{\sigma +\sigma }%
_{GS}^{M})^{-1}\partial _{g}\mathbf{d+}  \notag \\
&&\frac{1}{2}\text{Tr}\left[ [(\bm{\sigma +\sigma }_{GS}^{M})^{-1}%
\partial _{g}\bm{\sigma ]}^{2}\right] ,
\end{eqnarray}%
where $\mathbf{d}$ and $\bm{\sigma }$ are the displacement vector and
covariance matrix of the observed mode, respectively, $\bm{\sigma }%
_{GS}^{M}$ represents the covariance matrix of Gaussian measurement operator $\hat{\rho}_{GS}^{M}$. 
Notice that the first
moment $\mathbf{d}_{GS}^{M}$ of  $\hat{\rho}_{GS}^{M}$ does not affect $%
F_{g}^{GM}$ due to it is always possible to make $\mathbf{d}_{GS}^{M}=0$ by
the symplectic transformation before performing the measurement \cite{cenni2022thermometry}. By
comparing Eq.~(\ref{Eq31}) and Eq.~(\ref{Eq34}), one can easily obtain the measurement covariance
matrix corresponds to the Homodyne detection as   
\begin{equation}
\bm{\sigma }_{GS}^{M}=\bm{\sigma }_{\text{Ho}}^{M}=\lim_{r\rightarrow \zeta }\left[ 
\begin{array}{cc}
\frac{1}{r} & 0 \\ 
0 & r%
\end{array}%
\right] ,
\end{equation}%
where $\zeta =\infty $ and $0$ corresponds to the measurement of quadrature operators $\hat{X}$ and $\hat{Y}$, respectively. 
Similarly, for the Heterodyne detection, we have that 
\begin{equation}
\bm{\sigma }_{GS}^{M}=\bm{\sigma }_{\text{He}}^{M}=\left[ 
\begin{array}{cc}
1 & 0 \\ 
0 & 1%
\end{array}%
\right] .
\end{equation}

When  $\mathbf{d}$ and $\bm{\sigma }$ of the observation
mode are fixed, the optimal $\bm{\sigma }_{GS}^{M}$ (i.e., optimal Gaussian
measurement) must be found to maximize $F_{g}^{GM}$. However, this is in general a notoriously difficult task to analytically obtain the optimal Gaussian
measurement setup. The main obstacle stems from the objective function $F_{g}^{OGM}$
$=$ $\underset{\bm{\sigma }_{GS}^{M}}{\text{Max}}\{F_{g}^{GM}\}$ is a
nonlinear function of the measurement covariance matrix $\bm{\sigma }%
_{GS}^{M}$, hence $F_{g}^{OGM}$ can only be solved numerically. Note also that the quantum state to which the covariance matrix of the optimal Gaussian measurement belongs must be a pure state, which leads to $\bm{\sigma }_{GS}^{M}=\bm{%
	\sigma }_{\text{max}}^{M}\equiv S^{M}(S^{M})^{\text{T}}$ always holding \cite{serafini2023quantum}.
Here $S^{M}$ is a symplectic transformation, satisfying the symmetric constraint condition $S^{M}\Lambda (S^{M})^{%
	\text{T}}= \Lambda $. In this scenario, finding 
$F_{g}^{OGM}$ is transformed into the following semi-definite programming
(SDP) problem  \cite{cenni2022thermometry}: 
\begin{eqnarray}
F_{g}^{OGM} &:=&\underset{S^{M}}{\text{ Max}}\{F_{g}^{GM}(\mathbf{d},\bm{%
	\sigma };S^{M}(S^{M})^{\text{T}})\},  \notag \\
\text{s.t. }\Lambda  &= &S^{M}\Lambda (S^{M})^{\text{T}}.
\end{eqnarray}
Currently, we are still concerned with performing measurements on single-mode Gaussian states.

\section{Estimation of photon-magnon coupling strength} \label{IV}
In this section, we explore how various physical factors affect the estimation precision of the P-M coupling rate, such as the driving power $P_{l}$, the environment temperature $T$, the dissipation rates of cavity modes and magnon mode, the Kerr coefficient $K$, the photon tunneling rate $J$, the cavity modes detuning $ \Delta_{a}$, and the magnon detuning $ \Delta_{m}$. We then explore the optimal subsystem for estimating the P-M coupling parameter. Finally, the performance of Gaussian measurements performed on the optimal subsystem is evaluated.

For simplicity, we assume that the parameters of the two cavities are completely
consistent. Unless stated otherwise, here and in what follows, the parameters are chosen as: $P_{l}=500$ mW, $\omega _{l}=2\pi \times 10$ GHz, $%
T=10$ mK, $\gamma _{a_{1}}=\gamma _{a_{2}}=\gamma _{a}=2\pi \times 5$ MHz, $%
\gamma _{m}=2\pi \times 40$ MHz,  $\Delta _{a_{1}}=\Delta _{a_{2}}=\Delta
_{a}=2\pi \times 40$ MHz, $\Delta_{m}=2\pi \times 60$ MHz, $K=2\pi \times 2$ $\mu $Hz, $J=$ $2\pi \times 26$
MHz, $g= 2\pi \times 41$ MHz, whose values are mostly based on the latest experimental data \cite{bhoi2019photon,yuan2022quantum,rameshti2022cavity,zhang2019quantum,wang2018bistability,kong2019magnon}. From here we see that the cavity modes are indeed at microwave frequency band, which is necessary for the cavity mode coupled to the magnon via the magnetic-dipole interaction.  

The estimation precision reflects the sensitivity of the system to unknown parameters, whence the calculation of QFI and CFI involves derivative operations for the parameters to be estimated. As a technical remark, we point out that the first-order derivative of any $g$-dependent function $f_{g}$ with respect to $g$ during the numerical simulation is treated by the Lagrange interpolation method, i.e.,  \cite{sha2022continuous,zhang2022effects}
\begin{equation}
\frac{\partial f_{g}}{\partial g}:\simeq \frac{%
	-f_{g+2dg}+8f_{g+dg}-8f_{g-dg}+f_{g-2dg}}{12dg}.
\end{equation}%
This method of handling first-order derivatives has smaller errors and is more stable. Currently, we set $dg/g=10^{-6}$, which provides a very high accuracy. 

\subsection{Effect of power and  temperature} \label{IVA}
\begin{figure}[hbt!]
	\centering
	\includegraphics[width=0.95\linewidth]{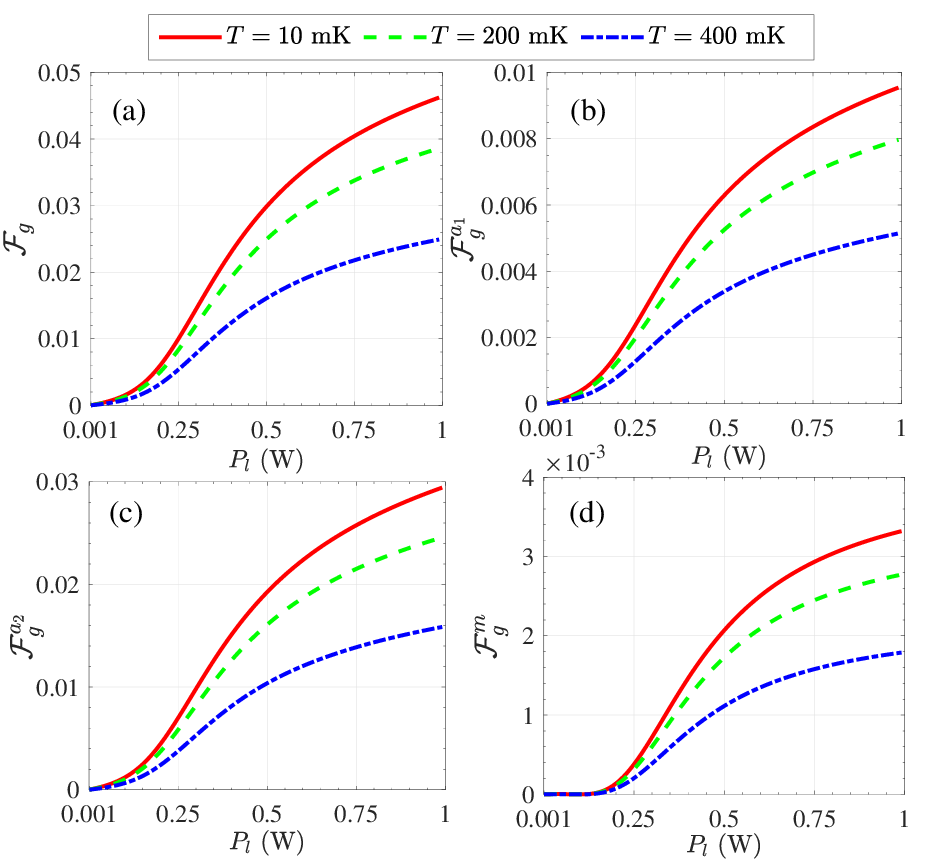}
	\caption{The QFIs versus the microwave driving power at different environment temperatures $T$, where QFI for (a) the global system, (b) the cavity mode $1$, (c) the cavity mode $2$, (d) the magnon mode.}
	\label{fig2}
\end{figure}
As shown in Fig.~\ref{fig2}, the QFIs for the whole system and the three subsystems are plotted as a function of microwave driving power $P_{l}$ at different environment temperatures, manifesting that all QFIs are nearly zero when the driving is weak ($P_{l}=1$ mW), i.e., the error for estimating the P-M coupling parameter is relatively large.
However, all QFIs are gradually enhanced with the increase of $P_{l}$, implying that microwave driving is beneficial for improving the estimation precision of $g$. 
This is easy to understand. Since the increase of external driving directly gives rise to an increase in the mean particle number of various modes in the double-cavity-magnon system, and the effective magnetic-dipole interaction is also enhanced, resulting in a reduced estimation error. 
Here, we emphasize that the mean particle number of all modes is much greater than $1$ even when $P_{l}=1$ mW, which ensures the linearization approximation holds. In addition, in the case of $P_{l}=1$ W, the mean magnon number $\left\langle \hat{m}^{\dag }\hat{m}\right\rangle \simeq 6\times 10^{13}\ll 2S=1.75\times 10^{17}$ holds for a 250-$\mu $m-diameter YIG sphere, indicating that the low-lying excitations assumption required for utilizing Holstein-Primakoff transformation in deriving the Hamiltonian $\hat{H}$ has not been violated (see Appendix \ref{AA}). 

Furthermore, one can see that the P-M coupling's information contained in the global state is always greater than that in the state of each subsystem.  
This also indirectly reflects the non-negative property of QFI, i.e., the more subsystems used, the higher QFI obtained.  On the other hand, the QFI of the global system is always greater than or equal to the independent summation of that of each subsystem. 
This is because the global system has some additional quantum correlation terms [see $C_{i,j}$ in Eq.~(\ref{eq19g})] compared to the direct-sum of the subsystems, where the quantum correlation terms also contain information about $g$. 
Particularly, the imprint of $g$ is mainly in cavity mode $2$, followed by cavity mode $1$, while the magnon mode contains the least, i.e., $\mathcal{F}_{g}^{a_{2}}>\mathcal{F}_{g}^{a_{1}}>\mathcal{F}_{g}^{m}$ in the given the parameter regime.  
Obviously, at this point the cavity mode $2$ is the optimal subsystem for estimating $g$. Physically, the distribution of the imprint of $g$ among subsystems relies on the interactions and correlations between subsystems. Later, we will specifically discuss this issue.

Consistent with our expectation, as the temperature increases (red$\rightarrow $green$\rightarrow $blue), all the QFIs drop off, indicating that adding thermal fluctuation to the system always decreases the estimation precision. Physically, the thermal fluctuations in general lead to a degradation of the quantum correlations of the system, thereby increasing the estimation error of P-M coupling rate.  
Note that the boosting effects of the quantum resources on estimation precision have been realized by many researchers working in the field.

\subsection{Effect of the damping channels} \label{IVB}
\begin{figure}[hbt!]
	\centering
	\includegraphics[width=0.95\linewidth]{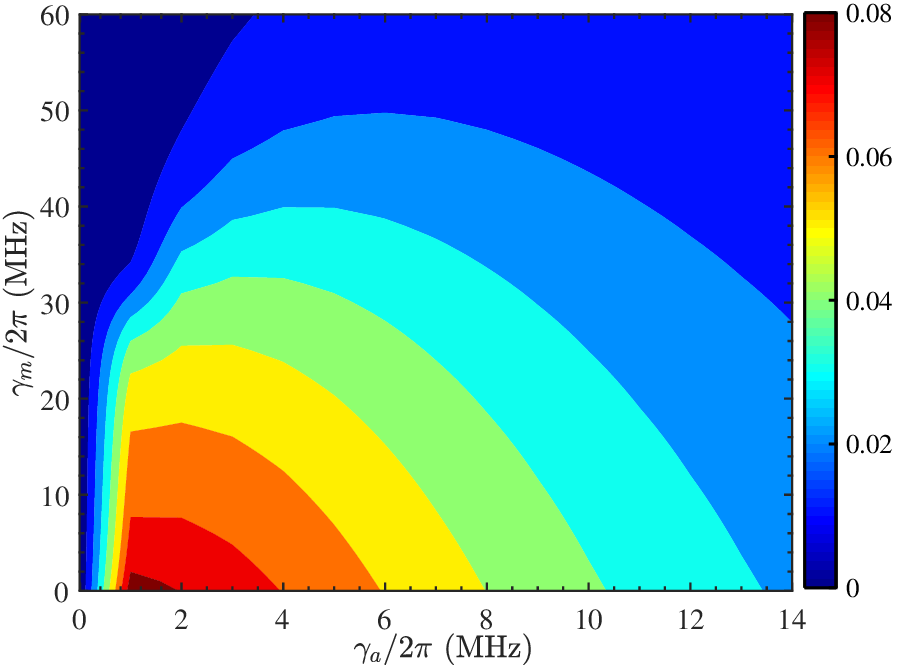}
	\caption{The density plot represents the QFI for the global system as a function of cavity mode loss $\gamma _{a}$ and magnon damping $\gamma _{m}$.}
	\label{fig3}
\end{figure}

It is of practical importance to investigate the influence of the dissipation rate on the estimation error. Presented in Fig.~\ref{fig3}, the density plot represents the global QFI $\mathcal{F}_{g}$ as a function of the cavity mode decay rate $\gamma _{a}$ and the magnon damping $\gamma _{m}$.  
According to Fig.~\ref{fig3}, it reveals that with the increase of $\gamma _{m}$ keeping $\gamma _{a}$ unaltered $\mathcal{F}_{g}$ monotonically decreases, i.e., the magnon dissipation is always detrimental for estimating the P-M coupling rate. 
In contrast, for a fixed $\gamma _{m}$ the estimation error first decreases and then increases with the increase of cavity decay rate $\gamma _{a}$. 
This indicates that there exists an optimized value of $\gamma_{a}$ where the estimation error reaches a minimum.

The physical reason behind this counterintuitive phenomenon is that: when $\sqrt{\gamma _{a}}$ is small, increasing $\gamma _{a}$ causes the increase of the microwave driving strength owing to $E_{l}\propto \sqrt{\gamma _{a}}$, resulting in enhancement of estimation precision (using the results of Fig.~\ref{fig2}).
Nevertheless, when $\gamma _{a}$ increases further, the dissipation of the cavity modes dominates the dynamics, which yields a significant increase in the estimation error. As a result, the final estimation precision is determined by the competition between the two opposite effects caused by the cavity mode loss.  The behavior of the QFI of each subsystem is rather similar to that of the global system, hence their figures are omitted here. 

In addition, we should also point out that if the loss rates of the two microwave cavities are different, when fixing the rate of cavity mode $1$, increasing that of cavity mode $2$ will obtain similar results. However, keeping the loss rate of cavity mode $2$ fixed while increasing that of cavity mode $1$ only leads to a decrease in estimation precision. This is due to the fact the driving field is applied on the cavity mode $2$ rather than the cavity mode $1$.

\subsection{Effect of Kerr nonlinearity and photon tunneling} \label{IVC}
\begin{figure}[hbt!]
	\centering
	\includegraphics[width=0.92\linewidth]{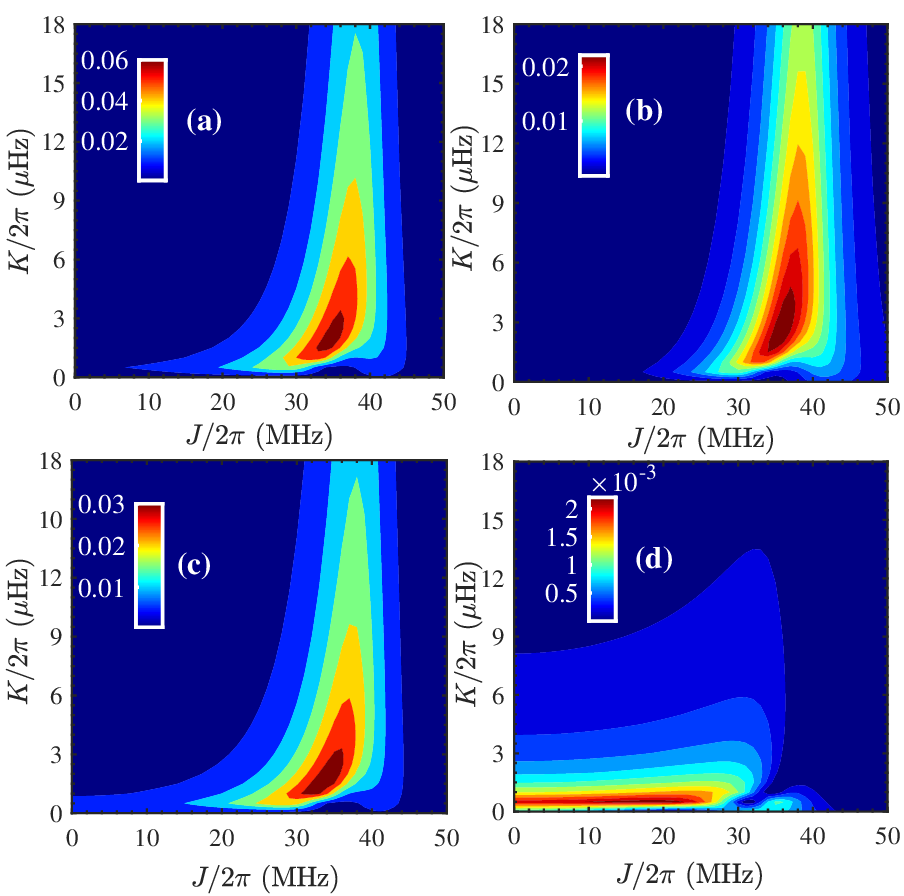}
	\caption{The QFIs vary with the Kerr coefficient $K$ and the photon tunneling rate $J$, where QFI for (a) the global system, (b) the cavity mode $1$, (c) the cavity mode $2$, (d) the magnon mode.}
	\label{fig4}
\end{figure}

In Fig.~\ref{fig4}, the density plot denotes the QFIs as a function of the Kerr coefficient $K$ and the photon tunneling rate $J$, manifesting that although the magnon mode is a necessary element to realize P-M coupling, most of the global QFI comes mainly from the contribution of the two cavity mode subsystems, with a small contribution from the magnon. 
One reason is that the damping of the magnon is greater than that of the cavity modes (the incoherence effect of the magnon modes is greater), while another reason is the photon tunneling induced cavity modes swap, i.e., dissemination of information about $g$. In this sense, cavity mode $1$ can act as an auxiliary mode to carry information about the P-M coupling rate, which may help improve estimation precision. 
Indeed, we can see that by appropriately designing the tunneling rate $J$, the global QFI increases, manifesting that the auxiliary cavity can reduce the estimation error in comparison with the unassisted case ($J=0$). 
This is because effective P-M coupling depends not only on the coupling strength, but is also closely related to the frequency matching between the modes, and photon tunneling interactions can adjust the frequencies of the mixed cavity modes.
Similar to the global QFI, the QFI of cavity mode $2$ is also significantly improved under appropriate photon tunneling rates. This is of great practical importance because, as we will see later, the cavity mode $2$ is the optimal subsystem for making Gaussian measurements in most parameter regions. 
Note that when the photon tunneling interaction is completely dominant (i.e., $J$ is sufficiently large), this is also detrimental to the estimation since the magnetic-dipole interaction can be almost neglected in the cavity-magnon system. 

On the other hand, all QFIs display a tendency to increase first and then decrease when fixed $J$ increases $K$. We can see that the appropriate Kerr coefficient can also greatly enhance the estimation precision compared to the case without the Kerr effect. Physically, this is due to the fact that the Kerr nonlinear effect causes quantum squeezing on the magnon.
In addition, the appropriate Kerr-effect-induced magnon frequency shift is also a reason for the improved estimation precision, as it can enhance effective P-M coupling. Nevertheless, as $K$ is further increased, the effective frequency difference between the magnon and the cavity modes is increased. This will weaken the effective P-M coupling interaction, increasing estimation error. The above analysis indicates that, when other parameters are fixed, a wise match between the Kerr nonlinearity coefficient and the photon tunneling rate is required to obtain the highest estimation precision. 

\subsection{Effect of detuning} \label{IVD}
\begin{figure}[hbt!]
	\centering
	\includegraphics[width=0.98\linewidth]{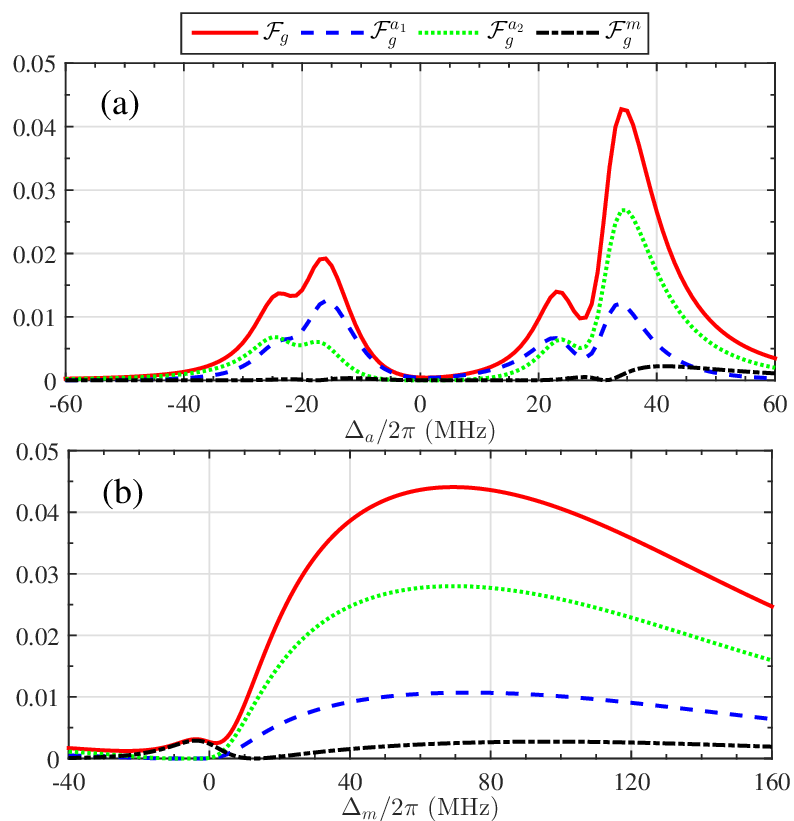}
	\caption{The QFIs of global system and subsystems as a function of  (a)  the cavity mode detuning $\Delta_{a}$ and  (b)  the magnon detuning $\Delta_{m}$, respectively. }
	\label{fig5}
\end{figure}

In Fig.~\ref{fig5}(a), we depict the QFIs as a function of the cavity mode detuning $\Delta_{a}$, showing that the global QFI still mainly comes from the two cavity modes. In particular, we can see that there are two peaks (one is the main peak and the other is the secondary peak) on each of the red, green, and blue curves in the red-detuned region. Similar results are also observed in the blue-detuned region, but the secondary peak is not significant. 
Physically, these valley-peak structures of QFIs originate from the presence of two hybridized cavity modes interacting with the magnon mode.
Particularly, the locations of these peaks are not only dependent on the selection of $\Delta_{m}$ but are also closely related to photon tunneling and the Kerr effect since both can adjust the effective mode detuning. The normal mode picture in Appendix \ref{AD} makes this clear.
In addition, we find that the red-detuned region is more conducive to estimating $g$, giving the highest estimation precision. This indicates that even including photon tunneling and Kerr nonlinear effects, in the red detuning region, the effective P-M coupling is still relatively strong due to $\Delta_{m}$ being greater than $0$. Note that this result holds without considering tunneling and Kerr nonlinearity effects. This is because the P-M coupling is essentially beam-splitter-like coupling, and the closer the frequencies of the two modes are, the stronger the coupling will be. 

Figure \ref{fig5}(b) plots the QFIs as a function of the magnon detuning $\Delta_{m}$. One can see that in the $\Delta _{m}<0$ region, all QFIs are relatively small, implying a higher estimation error. In the $\Delta _{m}>0$ region, all QFI curves exhibit a wide peak, indicating that we are easier to obtain high estimation precision. 
This is because the Kerr effect can induce an appreciable frequency shift of the magnon mode, resulting in a wide peak that is insensitive to $\Delta_{m}$ \cite{wang2016magnon}.  
Note that the frequency of the magnon can be flexibly adjusted by the external bias magnetic field $H_{B}$, which indirectly changes the $\Delta _{m}$, thus realizing the modulation of the estimation precision.
This means that when other parameters are fixed, there exists some ideal $H_{B}$ that can provide relatively high estimation precision.

\subsection{Optimal subsystem for estimating $g$} \label{IVE}
\begin{figure}[hbt!]
	\centering
	\includegraphics[width=0.9\linewidth]{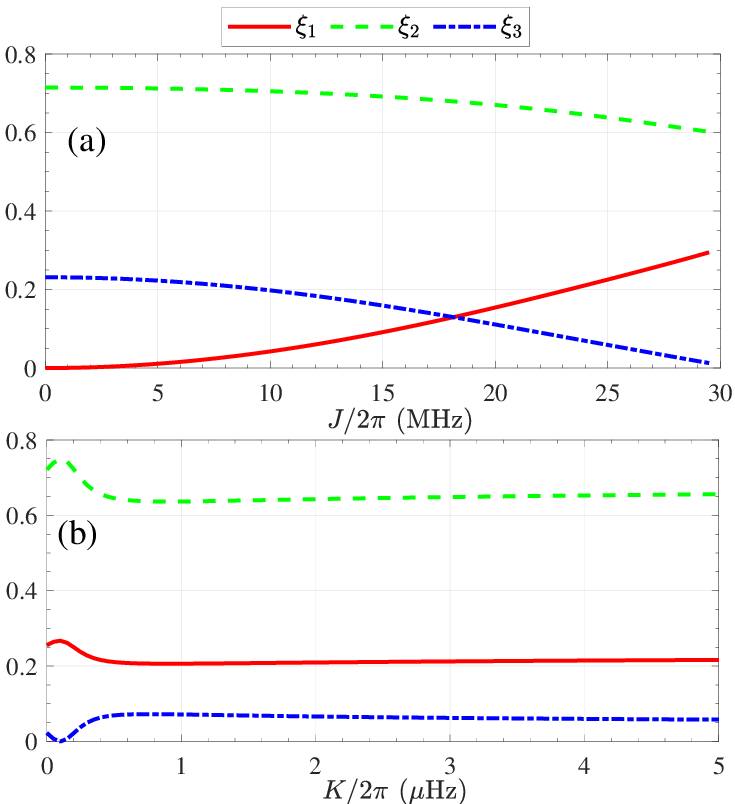}
	\caption{The ratio of the QFI of the subsystem to that of the global system is shown as a function of (a) the photon tunneling rate $J$ and (b) the Kerr coefficient $K$, respectively.}
	\label{fig6}
\end{figure}
Notice that, it is very difficult or even infeasible to measure the entire double-cavity-magnon system owing to the limitations of the measurement means. Thus a  wise choice would be to access and measure one of the subsystems. In this scenario, it becomes crucial to determine the optimal subsystem, i.e., the one that contains the most information about the estimated parameter. To this end, we define the ratio of the QFI of each subsystem to the global QFI, namely $\xi _{1}=\mathcal{F}_{g}^{a_{1}}/\mathcal{F}_{g}$, $\xi _{2}=\mathcal{F}_{g}^{a_{2}}/\mathcal{F}_{g}$, and $\xi _{3}=\mathcal{F}_{g}^{m}/\mathcal{F}_{g}$, respectively.

In the discussion about Fig.~\ref{fig2}, we have pointed out that cavity mode $2$ is the optimal subsystem for estimating the P-M coupling strength, indicating that $\xi _{2}$ is greater than $\xi _{1}$ and $\xi _{3}$. In addition, when studying the impact of damping channels on estimation precision, the result that cavity mode $2$ is the optimal subsystem still holds (note that in order to avoid figure duplication, the QFIs of the subsystems are not drawn in the Fig.~\ref{fig3}). However, according to Fig.~\ref{fig5}, we find that cavity mode $2$ is not always the optimal subsystem, as seen in the blue-detuned region. But in the red-detuned region, the cavity mode $2$ is an ideal candidate for extracting information from $g$. Below, we will focus on the impact of photon tunneling rate $J$ and Kerr nonlinearity coefficient $K$ on $\xi _{j}$ $(j=1,2,3)$.

Figure \ref{fig6}(a) plots $\xi _{j}$ $(j=1,2,3)$ as functions of the photon tunneling
strength $J,$ indicating that when $J=0$, $\xi _{2}>\xi _{3}>\xi _{1}=0$ holds. In other words, accessing cavity mode $2$ is able to pick up the most information about the P-M coupling parameter, followed by the magnon, while cavity mode $1$ does not contain information about $g$. This originates from the fact that no information can swap between cavity modes $1$ and $2$ without photon hopping interaction, so that $\xi _{1}=0$. The reason for $\xi _{2}>\xi _{3}$ is that the photon number of the cavity mode $2$ is more than the magnon number owing to the direct driving and $\gamma _{a}<\gamma _{m}$. With the increase of $J$, one find that  $\xi _{2}$ and$\ \xi _{3}$ decrease simultaneously, while $\xi _{1}$
increases. This reveals that part of the information about $g$ is transferred to cavity mode $1$ via the photon tunneling interaction. Moreover, the
stronger the photon tunneling effect, the greater the information containing $g$ in cavity mode $1$, resulting in  $\xi _{2}>\xi _{1}>\xi _{3}$.
In particular, $\xi _{2}>\xi _{1},\xi _{3}$ holds all the time, manifesting that the amount of information about the P-M coupling rate in cavity mode $2$ is always the most, i.e., the cavity mode $2$ is the optimal subsystem for estimating $g$ for the given parameter regime.

In Fig.~\ref{fig6}(b), we present behaviors of $\xi _{j}$ $(j=1,2,3)$ versus the Kerr coefficient $K$. 
One can see that when $K$ starts increasing from $0$, $\xi_1$ and $\xi_2$ increase first and then decrease, while $\xi_3$ has the opposite trend. 
This is due to the fact that the trade-off and competition between the P-M coupling and the photon tunneling can be affected by the Kerr self-interaction of magnon, while the first two play a decisive role in the exchange of information between subsystems.
With a further increase in $K$, the self-interaction of magnon completely surpasses photon tunneling and P-M coupling, so that the distribution of information in each subsystem no longer changes significantly.
Indeed, when changing $J$ or $K$, the behavior of $\xi _{3}$ is always opposite to that of the other two, which reflects information swap very well. 
Importantly, $\xi _{2}>\xi _{1},\xi_{3}$ holds all the time, i.e., the cavity mode $2$ is always the optimal
subsystem for estimating the P-M coupling rate.

\subsection{Performance analysis of Gaussian measurements} \label{IVF}
According to subsection \ref{IVE}, we already know that cavity mode $2$ is the optimal subsystem for estimating $g$  in most parameter regions except in the blue-detuned region where the QFI is small. This means that accessing cavity mode $2$ for obtaining information about the P-M coupling parameter is relatively high efficiency in most cases.
Consequently, considering that only one subsystem can be accessed, we only focus on the case of the information about $g$ obtained by performing measurements on the cavity mode $2$. 

\begin{figure}[hbt!]
	\centering
	\includegraphics[width=0.95\linewidth]{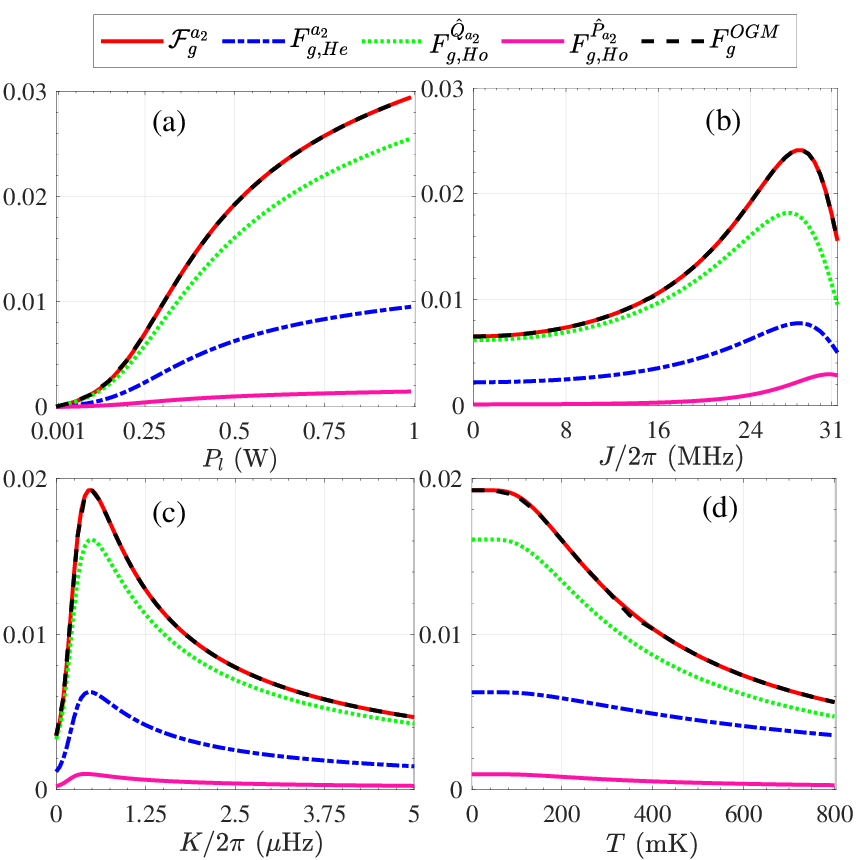}
	\caption{The QFI $\mathcal{F}_{g}^{a_{2}}$ and CFIs for the cavity mode 2
		against (a) the microwave driving power $P_{l}$, (b) the photon tunneling
		strength $J$, (c) the Kerr coefficient $K$, (d) the ambient temperature $T$,
		where $F_{g,\text{He}}^{a_{2}}$, $F_{g,\text{Ho}}^{\hat{Q}_{a_{2}}}$, $F_{g,\text{Ho}}^{\hat{P}_{a_{2}}}$
		and  $F_{g}^{\text{OGM}}$ corresponding to Heterodyne detection, Homodyne detection for $\hat{Q%
		}_{a_{2}}$ and $\hat{P}_{a_{2}}$ quadrature operators, and optimal Gaussian
		measurement, respectively.}
	\label{fig7}
\end{figure}

In Fig.~\ref{fig7}, we compare the
ultimate precision bound given by QFI $\mathcal{F}_{g}^{a_{2}}$ with the precisions achieved through the Gaussian measurements. One find that $\mathcal{F}_{g}^{a_{2}}>F_{g,\text{Ho}}^{\hat{Q}%
	_{a_{2}}}>F_{g,\text{He}}>F_{g,\text{Ho}}^{\hat{P}_{a_{2}}}$ always holds, indicating that
neither widely available Homodyne detection nor Heterodyne detection is the optimal measurement setup. In addition, the Homodyne measurement for amplitude quadrature $\hat{Q}%
_{a_{2}}$  is better than the Heterodyne detection, and the worst scheme
is the Homodyne detection for phase quadrature $\hat{P}_{a_{2}}$. Roughly speaking, measuring the $\hat{P}_{a_{2}}$ quadrature is hardly helpful for estimating the P-M coupling because the obtained CFI is too small. Physically, this is because the P-M coupling mainly affects the amplitude of cavity mode $2$ (i.e., mean photon number of cavity mode 2)  rather than the phase, so the amplitude quadrature of cavity mode $2$ contains more information about $g$. In particular,
we also obtain the CFIs corresponding to the optimal Gaussian measurement by means of SDP (see black dashed lines).  
We find that the red solid line and the black dashed line almost overlap in all subfigures, i.e., the optimal Gaussian measurement almost constitutes the optimal setup for estimating the P-M coupling strength.

It should be pointed out that the optimal POVM measurement setup is constituted by a set of projection operators over the eigenvectors of symmetric logarithmic derivative $L_{g}^{a_{2}}$, where $L_{g}^{a_{2}}$ is composed of the first moments and covariance matrix of the cavity mode $2$ \cite{bakmou2020multiparameter,pinel2012ultimate,monras2013phase,vsafranek2018estimation}. In principle, one can always theoretically construct optimal measurements based on this conclusion. However, such optimal measurement is generally not experimentally feasible, especially in the current model where such interactions are relatively complex. In this scenario, the optimal Gaussian measurement is much more favored because it is experimentally feasible. In addition, the practical imperfections tend to offset the difference between optimal and nearly optimal setups in the laboratory \cite{oh2019optimal}. 
This also reflects the fact that the estimation precision limit given by $\mathcal{F}_{g}^{a_{2}}$ is experimentally achievable. Unfortunately, due to difficulties in mathematical techniques, we do not yet know the specific form of the optimal Gaussian measurement. In theory, it should be the product of the quadrature operators of the observed mode (e.g., complex combination of amplitude and phase quadrature operators).
In addition, it may depend on the values of the parameters to be evaluated. In this sense, this optimal Gaussian measurement needs to be implemented adaptively by accumulated data. We will further investigate such a question elsewhere, which has enormous practical significance.

\section{Conclusion} \label{V}
In summary, we have explored the quantum parameter estimation problem of the P-M coupling strength in a driven-dissipative double-cavity-magnon system, where one cavity is primary and the other is auxiliary. 
We found that (i) with the increase of the driving power, the estimation error gradually decreases; (ii) temperature and magnon damping are always detrimental for estimating the P-M coupling strength; (iii) the dissipation rate of the cavity mode does not always have a detrimental effect on the estimation error, and the existence of a critical dissipation rate giving the highest estimation precision;  (iv) by designing appropriate Kerr coefficient and photon tunneling rate, the estimation error can be significantly reduced; (v) compared to the blue-detuned region, the red-detuned region of the cavities is more conducive to achieving high precision estimation. Moreover, the external bias magnetic field applied to the YIG sphere can indirectly adjust estimation error due to its ability to modulate magnon detuning; (vi) the summation of
QFI for all subsystems is always less than or equal to the QFI of the global system. Our analysis also revealed that the optimal subsystem for carrying out measurements and estimations is the primary cavity mode since in most parameter regions the majority of information about the P-M coupling rate is encoded in its reduced state.

Further, we explored the CFI obtained by performing different Gaussian measurements on the primary cavity mode and compared them with QFI to evaluate the practical performance of measurements. The results indicated that Homodyne and Heterodyne detections are not optimal measurement strategies for the extraction of the P-M coupling information. The Homodyne detection for amplitude quadrature surpasses the Heterodyne detection, and the worst is the Homodyne detection for phase quadrature. Particularly, the optimal Gaussian measurement is almost the optimal measurement strategy, i.e., it can extract almost all the information about the P-M coupling parameter from the primary cavity mode. 
We note that increasing the number of auxiliary cavities can improve the QFI, but at the same time introduces more dissipations and makes the experiment more difficult.
We also note that once the P-M coupling strength is comparable to the magnon or cavity mode frequencies, one can no longer consider photon and magnon dissipation separately, as is done here.
Potentially, our method could be used to investigate the estimation problem of other parameters in cavity-magnon systems, such as photon tunneling rate, frequency and Kerr coefficient of magnons. However, the optimal structure of the composite system may not be a double-cavity structure with one primary and one auxiliary. We believe that this work provides some insights into the use of magnetic-dipole interaction for quantum precision measurements or quantum information processing.

\section*{ACKNOWLEDGMENTS} \label{VI}
This work was supported by the Innovation Program for
Quantum Science and Technology (No. 2021ZD0303200); the
National Key Research and Development Program of China
(No. 2016YFA0302001); the National Science Foundation of
China (Nos. 12374328, 11974116, 12234014, and 11654005); the
Shanghai Municipal Science and Technology Major Project
(No. 2019SHZDZX01); the Fundamental Research Funds for
the Central Universities; the Chinese National Youth Talent
Support Program, and the Shanghai talent program.

\appendix
\section{Derivation of Hamiltonian for YIG sphere and cavity mode 2} \label{AA}
The quantized interaction form between cavity modes (coupled cavity) is well-known to everyone \cite{aspelmeyer2014cavity}.  We thus will only provide a detailed introduction to the interaction between the YIG sphere and cavity mode $2$. The cavity mode $2$ includes electric field energy and magnetic field energy, while the magnetized YIG sphere includes Zeeman energy, demagnetization energy and anisotropic energy of magnetocrystalline, and there is magnetic-dipole interaction between the two. Therefore, the total Hamiltonian  can be written as \cite{yuan2022quantum,rameshti2022cavity,blundell2001magnetism}
\begin{eqnarray}
\label{EA1}
\hat{H}_{\text{a}_{2}\text{m}} &=&\frac{1}{2}\int \left( \varepsilon _{0}%
\mathbf{E}^{2}+\frac{\mathbf{B}^{2}}{\mu _{0}}\right) d\mathbf{v}-\int 
\mathbf{M}\cdot \mathbf{H}_{B}d\mathbf{v}  \notag \\
&&-\frac{\mu _{0}}{2}\int \mathbf{M}\cdot \mathbf{H}_{\text{an}}d\mathbf{v}%
-\mu _{0}\int \mathbf{M}\cdot \mathbf{B}d\mathbf{v},
\end{eqnarray}
where $\mathbf{E}$ and $\mathbf{B}$ are the electric and magnetic components
of the electromagnetic field inside the cavity $2$, respectively; $\varepsilon
_{0}$ and $\mu _{0}$ are, respectively, the vacuum permittivity and vacuum
permeability; $\mathbf{H}_{B}=H_{B}$$\vec{\textbf{e}}_{\text{z}}$ being the stable
magnetic field applied along the Z-axis on the YIG sphere, aimed at
magnetizing the YIG sphere. The corresponding magnetization strength is $%
\mathbf{M} = \gamma _{e}$ \textbf{S}$/V_{m} = \mathbf{(}$M$_{\text{x}}, $M$_{\text{y}}, $M$_{\text{z}}%
\mathbf{)}$, in which \textbf{S}  $\equiv (S_{\text{x}},S_{\text{y}},S_{\text{z}})
$ denotes the macrospin and $V_{m}$ the volume of
the YIG sphere; $\mathbf{H}_{\text{an}}=(-2K_{\text{an}}$M$_{\text{z}}/$M$_{b}^{2})$%
$\vec{\textbf{e}}_{\text{z}}$ represents the anisotropic field owing to the
magnetocrystalline anisotropy (relying on the angle between the crystallographic axis of the YIG sphere and  the direction   of  the externally applied stable magnetic field)\cite{Soykal}, in which $K_{\text{an}}$
and M$_{b}$ are the first-order anisotropy constant of   the YIG sphere and the saturation
magnetization, respectively; the last term represents the magnetic-dipole interaction between cavity mode $2$ and magnon. Note that demagnetization energy is ignored in Eq.~(\ref{EA1}) because it is a constant term \cite{yuan2022quantum,rameshti2022cavity}.

Suppose that the magnetic field
direction inside the cavity $2$ is along the X-axis, i.e., $\mathbf{B=-}%
\sqrt{\omega _{a_{2}}/\mu _{0}V_{a}}(\hat{a}_{2}+\hat{a}_{2}^{\dag })$%
$\vec{\textbf{e}}_{\text{x}}$, where $V_{a}$ is the volume of the cavity 2.
One can introduce magnon mode to represent a collective excitation of a large number of spins by the Holstein-Primakoff transform \cite{Holstein}
\begin{subequations}
	\begin{eqnarray}
	\hat{S}_{\text{z}} &=&S-\hat{m}^{\dag }\hat{m}, \\
	\hat{S}_{+} &=&\sqrt{2S-\hat{m}^{\dag }\hat{m}}\hat{m}, \\
	\hat{S}_{-} &=&\hat{m}^{\dag }\sqrt{2S-\hat{m}^{\dag }\hat{m}}, \\
	\hat{S}_{\pm } &=&\hat{S}_{x}\pm i\hat{S}_{y},
	\end{eqnarray}%
\end{subequations}
where $S$ is the total spin number of the YIG sphere.  
In particular, $2S\gg
\langle \hat{m}^{\dag }\hat{m}\rangle $ in general holds owing to a fact $%
2S=5\rho V_{m}$ is very huge for a YIG sphere with spin density $\rho
\approx 4.22\times 10^{27}$ m$^{-3}$ \cite{bhoi2019photon,yuan2022quantum,rameshti2022cavity}. 
This results in the following is approximately valid: $\hat{S}_{+}\simeq \sqrt{2S}\hat{m}$ and $\hat{S}_{-}\simeq \sqrt{2S}%
\hat{m}^{\dag }$. By putting the pieces together, the $\hat{H}_{\text{a}_{2}%
	\text{m}}$ can be rewritten as
\begin{eqnarray}
\hat{H}_{\text{a}_{2}\text{m}}&=&\omega _{a_{2}}\hat{a}_{2}^{\dag }\hat{%
	a}_{2}-\gamma _{e}H_{B}\hat{S}_{\text{z}}+\frac{\mu _{0}K_{\text{an}}\gamma _{e}^{2}%
	\hat{S}_{\text{z}}^{2}}{V_{m}\text{M}_{b}^{2}}  \notag \\
&&+g_{am}(\hat{a}_{2}+\hat{a}_{2}^{\dag })(\hat{S}_{+}+\hat{S}_{-}),
\end{eqnarray}%
where $g_{am}=\sqrt{\mu _{0}\gamma _{e}^{2}\omega _{a_{2}}/4V_{a}}$ is the coupling strength between the cavity mode $2$ and the spins. 
Using $\sqrt{2S}\hat{m}$ and $\sqrt{2S}\hat{m}^{\dag }$ to replace $\hat{S}_{\pm }$, and considering
the rotating-wave approximation, one can finally obtain
\begin{eqnarray}
\hat{H}_{\text{a}_{2}\text{m}}&=&\omega _{a_{2}}\hat{a}_{2}^{\dag }\hat{%
	a}_{2}+\omega _{m}\hat{m}^{\dag }\hat{m}+K\hat{m}^{\dag }\hat{m}\hat{m}%
^{\dag }\hat{m}  \notag \\
&&+g(\hat{a}_{2}\hat{m}^{\dag }+\hat{a}_{2}^{\dag }\hat{m}),
\end{eqnarray}%
where $\omega _{m}=\gamma _{e}H_{B}-2\mu _{0}K_{\text{an}}\gamma
_{e}^{2}S/V_{m}$M$_{b}^{2}$ being  the frequency of the magnon, indicating that the
 frequency of the magnon  can be tuned by the bias magnetic field ; $K=\mu
_{0}K_{\text{an}}\gamma _{e}^{2}/V_{m}$M$_{b}^{2}$ is Kerr nonlinear
coefficient; $g=\sqrt{2S}g_{am}$ stands for the P-M coupling
strength.
\section{Stability conditions for the system} \label{AB}
The stability of the system is ensured by Routh-Hurwitz criterion \cite{dejesus1987routh}, namely
the all eigenvalues of the drift matrix $\mathfrak{A}$ have negative real parts, indicating
that the system is stable. To this end, we need to evaluate the
characteristic equation of $\mathfrak{A}$, i.e.,
$|\mathfrak{A}-\lambda \mathbb{1}_{6}|=0$, yielding the characteristic equation
\begin{equation}
\lambda ^{6}+\alpha _{1}\lambda ^{5}+\alpha _{2}\lambda ^{4}+\alpha
_{3}\lambda ^{3}+\alpha _{4}\lambda ^{2}+\alpha _{5}\lambda +\alpha _{6}=0,
\end{equation}%
where
\begin{eqnarray}
\alpha _{0}&=&1 \\
\alpha _{1} &=&4\gamma _{a}-\eta _{1}, \\
\alpha _{2} &=&2(g^{2}+J^{2})+6\gamma _{a}^{2}+2\Delta _{a}^{2}-4\eta
_{1}+\eta _{2}, \\
\alpha _{3} &=&4\gamma _{a}^{3}-\eta _{1}(g^{2}+2\eta _{5})-6\gamma
_{a}^{2}\eta _{1}+\gamma _{a}\mu _{0}, \\
\alpha _{4} &=&\gamma _{a}^{4}-4\gamma _{a}^{3}\eta _{1}+\gamma _{a}^{2}\mu
_{2}-\gamma _{a}\mu _{3}+2\Re _{+}\Re _{-}\eta _{5}, \\
\alpha _{5} &=&2g^{4}\gamma _{a}-J^{4}\eta _{1}+g^{2}\mu _{5}-2J^{2}\mu
_{6}-\eta _{4}\mu _{7}, \\
\alpha _{6} &=&g^{4}\eta _{4}+\eta _{2}\mu _{8}-g^{2}\mu _{9,}
\end{eqnarray}%
with
\begin{eqnarray*}
	\eta _{1} &=&\Re _{+}+\Re _{-},\eta _{2}=\Re _{+}\Re _{-}-\Im _{+}\Im _{-},
	\\
	\eta _{3} &=&\Im _{+}-\Im _{-},\eta _{4}=\gamma _{a}^{2}+\Delta _{a}^{2}, \\
	\eta _{5} &=&J^{2}+\Delta _{a}^{2},\mu _{0}=6g^{2}+4(\eta _{5}+\eta _{2}), \\
	\mu _{1} &=&g^{4}+2g^{2}J^{2}+J^{4}-2J^{2}\Delta _{a}^{2}+\Delta _{a}^{4}, \\
	\mu _{2} &=&6\left( g^{2}+\eta _{2}\right) +2\eta _{5},\mu _{3}=(4\Delta
	_{a}^{2}+3g^{2}+4J^{4})\eta _{1}, \\
	\mu _{4} &=&2\Im _{+}\Im _{-}\eta _{5}-g^{2}\Delta _{a}(2\Delta _{a}-\eta
	_{3}), \\
	\mu _{5} &=&2\gamma _{a}^{3}+J^{2}(2\gamma _{a}-\eta _{1})-(3\gamma
	_{a}^{2}+\Delta _{a}^{2}) \\
	&&\eta _{1}+2\gamma _{a}\Delta _{a}(\Delta _{a}-\Im _{+}+\Im _{-}), \\
	\mu _{6} &=&\eta _{4}\eta _{1}-2\gamma _{a}\eta _{2},\mu _{7}=\eta _{4}\eta
	_{1}-4\gamma _{a}\eta _{2}, \\
	\mu _{8} &=&J^{4}+2J^{2}(\gamma _{a}^{2}-\Delta _{a}^{2})+\eta _{4}^{2}, \\
	\mu _{9} &=&\eta _{4}(\gamma _{a}\eta _{1}+\Delta _{a}\eta
	_{3})+J^{2}(\gamma _{a}\eta _{1}-\Delta _{a}\eta _{3}).
\end{eqnarray*}
Note that for simplicity, here we have assumed that the parameters of the two cavity modes are completely consistent, i.e., $\gamma _{a_{1}}=\gamma _{a_{2}}=\gamma _{a}$ and $\Delta _{a_{1}}=\Delta _{a_{2}}=\Delta
_{a}$ (All numerical results in the main text are also based on this assumption). 
Based on coefficient $\alpha _{k}$, one can construct $6$ Hurwitz matrices,
where the dimension of the $k$-th matrix is $k\times k$ $\left( 1\leq k\leq
6\right) $, the corresponding matrix elements are determined by the
following conditions \cite{Jiao}
\begin{equation}
\label{EqA8}
\mathfrak{H}_{ij}^{k}=
\begin{cases}
0,& 2i-j<0\text{ or}\ 2i-j>k,\\
\alpha_{2i-j},& \text{otherwise},
\end{cases}
\end{equation}
where $1\leq i,j\leq k$. For example, when $k=1$ and $3$, we can easily obtain by Eq.~(\ref{EqA8})
\begin{equation}
\mathfrak{H}^{1}=\left[ \alpha _{1}\right],\mathfrak{H}^{3}=\left[ 
\begin{array}{ccc}
\alpha _{1} & 1 & 0 \\ 
\alpha _{3} & \alpha _{2} & \alpha _{1} \\ 
0 & 0 & \alpha _{3}%
\end{array}%
\right] .
\end{equation}
The stability condition of the system is that all the determinants of Hurwitz matrices are positive, i.e.,
\begin{equation}
\label{EqA10}
\forall \ \text{det}\left[ \mathfrak{H}^{k}\right] >0 \ \text{holds},1\leq k\leq 6.
\end{equation}
Base on Eq.~(\ref{EqA10}), the following conditions are obtained, i.e., 
\begin{eqnarray}
\label{EqA11}
\alpha _{i} &>&0\text{ }(1\leq i,j\leq k);\alpha _{1}\alpha _{2}>\alpha
_{3}, \\
\alpha _{1}\alpha _{2}\alpha _{3} &>&\alpha _{3}^{2}+\alpha _{1}^{2}\alpha
_{4};T_{1}>T_{2};T_{3}>T_{4},\label{EqA12}
\end{eqnarray}%
with
\begin{eqnarray*}
	T_{1} &=&(\alpha _{1}\alpha _{4}-\alpha _{5})(\alpha _{1}\alpha _{2}\alpha
	_{3}-\alpha _{3}^{2}-\alpha _{1}^{2}\alpha _{4}), \\
	T_{2} &=&\alpha _{1}\alpha _{5}^{2}+\alpha _{5}(\alpha _{1}\alpha
	_{2}-\alpha _{3})^{2}, \\
	T_{3} &=&\alpha _{1}^{2}\alpha _{6}(2\alpha _{2}\alpha _{5}+\alpha
	_{3}\alpha _{4})+\alpha _{3}^{3}\alpha _{6}+ \\
	&&\alpha _{1}\alpha _{2}\alpha _{3}\alpha _{4}\alpha _{5}+\alpha
	_{5}^{2}(2\alpha _{1}\alpha _{4}+\alpha _{2}\alpha _{3}) \\
	T_{4} &=&\alpha _{1}^{2}(\alpha _{1}\alpha _{6}^{2}+\alpha _{4}^{2}\alpha
	_{5})+\alpha _{5}^{3}+\alpha _{4}\alpha _{5}\alpha _{3}^{2} \\
	&&+\alpha _{1}(\alpha _{2}\alpha _{6}\alpha _{3}^{2}+3\alpha _{3}\alpha
	_{5}\alpha _{6}+\alpha _{2}^{2}\alpha _{5}^{2}).
\end{eqnarray*}
 The Eqs.~(\ref{EqA11})-(\ref{EqA12}) ensure the stability of the driven-dissipative double-cavity-magnon system. 
\\

\section{Normal mode picture} \label{AD}
The effective Hamiltonian of the double-cavity-magnon system is 
\begin{eqnarray}
\label{BB1}
\hat{H}_{\text{eff}} &=&\Delta _{1}\delta \hat{a}_{1}^{\dag }\delta \hat{a}%
_{1}+\Delta _{2}\delta \hat{a}_{2}^{\dag }\delta \hat{a}_{2}+\Delta
_{\text{eff}}\delta \hat{m}^{\dag }\delta \hat{m}  \notag \\
&&+K\left[\mathbf{\langle }\hat{m}\mathbf{\rangle }^{2}\delta \hat{m}^{\dag
}\delta \hat{m}^{\dag }+\mathbf{\langle }\hat{m}\mathbf{\rangle }^{\ast
	^{2}}\delta \hat{m}\delta \hat{m}\right]\ + \\
&&J(\delta \hat{a}_{1}^{\dag }\delta \hat{a}_{2}+\delta \hat{a}_{1}\delta 
\hat{a}_{2}^{\dag })+g(\delta \hat{a}_{2}^{\dag }\delta \hat{m}+\delta \hat{a%
}_{2}\delta \hat{m}^{\dag }).  \notag
\end{eqnarray}%
Based on the above equation, one can clearly see that only cavity mode 2 and the magnon are coupled
via beam-splitter-like interaction while the magnon is subject to single-mode squeezing. 
However, in the normal mode picture, the two interactions take on a different form.

Introducing  Bogoliubov  transformation \cite{fetter2012quantum}
\begin{subequations}
\label{BB2}
\begin{eqnarray}
 \hat{M} &:&=\alpha \delta \hat{m}-\beta ^{\ast }\delta \hat{m}^{\dag }, \\
 \hat{M}^{\dag } &:&=\alpha ^{\ast }\delta \hat{m}^{\dag }-\beta \delta \hat{%
	m},
\end{eqnarray}%
\end{subequations}
with
\begin{eqnarray}
\alpha  &=&\sqrt{(\Delta _{\text{eff}}/\mathcal{E}+1)/2},\beta e^{i\phi }=-\sqrt{(\Delta _{\text{eff}}/\mathcal{E}-1)/2},%
 \notag \\
\mathcal{E} &=&\sqrt{\Delta _{\text{eff}}^{2}-4|\mathbf{\langle }\hat{m}\mathbf{\rangle }%
	|^{4}K^{2}},\phi =\text{arctan}(\mathcal{I}/\mathcal{R}), \notag
\end{eqnarray}
where $\mathcal{I}=$ Im$(2K\mathbf{\langle }\hat{m}\mathbf{\rangle }^{2})$ and $\mathcal{R}=$ Re$(%
2K\mathbf{\langle }\hat{m}\mathbf{\rangle }^{2})$ quantified  the magnetocrystalline anisotropy of YIG sphere.
Substituting Eqs.~(\ref{BB2}) into Eq.~(\ref{BB1}), $\hat{H%
}_{\text{eff}}$ can be written as
\begin{eqnarray}
\label{EqC33}
\hat{\mathcal{H}}_{\text{eff}} &=&\Delta _{1}\delta \hat{a}_{1}^{\dag }\delta \hat{a}%
_{1}+\Delta _{2}\delta \hat{a}_{2}^{\dag }\delta \hat{a}_{2}+\mathcal{E}\hat{M}^{\dag
} \hat{M}  \notag \\
&&+g\left[ (\beta  \hat{M}+\alpha  \hat{M}^{\dag })\delta \hat{a}_{2}+(\beta
^{\ast }\hat{M}^{\dag }+\alpha ^{\ast }\hat{M})\delta \hat{a}_{2}^{\dag }%
\right]   \notag \\
&&+J(\delta \hat{a}_{1}^{\dag }\delta \hat{a}_{2}+\delta \hat{a}_{1}\delta 
\hat{a}_{2}^{\dag }),
\end{eqnarray}%
where $g(\beta \hat{M}\delta \hat{a}_{2}+\beta ^{\ast }\hat{M}^{\dag
}\delta \hat{a}_{2}^{\dag })$ being the squeezing-like coupling, resulting in
the entanglement between the normal magnon mode $\hat M$ and the cavity mode 2. Notice also that
Kerr coefficient $K=0$ makes $\beta =0$ owing to $\Delta _{\text{eff}}=\mathcal{E}$, leading to the squeezing-like coupling disappear. This indicates that the magnetocrystalline anisotropy is the key to inducing the entanglement between the magnon and the cavity mode 2.

In order to clearly show the two cavity-magnon interactions corresponding to the double-peak structure in Fig.~\ref{fig5}(a), we further introduce the following transformation, i.e.,
\begin{subequations}
\begin{eqnarray}
\hat{A}_{+} &:=&f\delta \hat{a}_{1}-h\delta \hat{a}_{2}, \\
\hat{A}_{-} &:=&h\delta \hat{a}_{1}+f\delta \hat{a}_{2}.
\end{eqnarray}%
\end{subequations}
Substitute the above formula into the Eq.~(\ref{BB1}), we can obtain 
\begin{eqnarray}
\hat{H}_{\text{eff}} &=&\Delta _{\text{eff}}\delta \hat{m}^{\dag }\delta \hat{m}+K\left[\mathbf{%
	\langle }\hat{m}\mathbf{\rangle }^{2}\delta \hat{m}^{\dag }\delta \hat{m}%
^{\dag }+\mathbf{\langle }\hat{m}\mathbf{\rangle }^{\ast ^{2}}\delta \hat{m}%
\delta \hat{m}\right]  \notag \\
&&+ \omega_{+}\hat{A}_{+}^{\dag }\hat{A}_{+}+\omega_{-}%
\hat{A}_{-}^{\dag }\hat{A}_{-}+G_{+}\left( \delta \hat{m}\hat{A%
}_{+}^{\dag }+\delta \hat{m}^{\dag }\hat{A}_{+}\right)   \notag \\
&&+G_{-}\left( \delta \hat{m}\hat{A}_{-}^{\dag }+\delta \hat{m}%
^{\dag }\hat{A}_{-}\right) ,
\end{eqnarray}%
where $\omega_{\pm }$ refers to the resonance frequency of hybridized cavity modes; $G_{\pm }$ is the coupling strengths between the hybridized cavity modes and the magnons. Their specific forms are
\begin{eqnarray}
G_{+} &=&-gh,G_{-}=fg, \\
\omega_{\pm } &=&\frac{1}{2}\left[\left( \Delta _{1}+\Delta _{2}\right)
\pm \sqrt{\left( \Delta _{1}-\Delta _{2}\right) ^{2}+4J^{2}}\right],
\end{eqnarray}%
with $f=\left\vert \omega_{-}-\Delta _{1}\right\vert /\sqrt{\left( 
	\omega_{-}-\Delta _{1}\right) ^{2}+J^{2}}$ and $h=Jf/\left(\omega
_{-}-\Delta _{1}\right)$, in which $f^{2}+h^{2}=1$.

Further, Eq.~\ref{EqC33} can be rewritten as
\begin{eqnarray}
\label{EqD5}
\hat{\mathcal{H}}_{\text{eff}} &=&\omega _{+}\hat{A}_{+}^{\dag }\hat{A}_{+}+\omega
_{-}\hat{A}_{-}^{\dag }\hat{A}_{-}+\mathcal{E}\hat{M}^{\dag }
\hat{M} \\
&&+G_{+}\left[ \left( \beta \hat{M}+\alpha \hat{M}^{\dag }\right) \hat{A}%
_{+}+(\beta ^{\ast }\hat{M}^{\dag }+\alpha ^{\ast }\hat{M})\hat{A}_{+}^{\dag
}\right]   \notag \\
&&+G_{-}\left[ \left( \beta \hat{M}+\alpha \hat{M}^{\dag }\right) \hat{A}%
_{-}+(\beta ^{\ast }\hat{M}^{\dag }+\alpha ^{\ast }\hat{M})\hat{A}_{-}^{\dag
}\right].  \notag
\end{eqnarray}
From  Eq.~(\ref{EqD5}), we can clearly see that the interaction between the magnon and the two cavity modes is essentially equivalent to the that between the normal magnon mode $\hat{M}$ and the two hybrid cavity modes ($\hat{A}_{+} $ and $\hat{A}_{-} $).
\section*{References}
\bibliography{Ref}

\end{document}